\documentclass[12pt]{JHEP3}
\usepackage{amsmath,amsfonts,amsbsy}
\usepackage[dvips]{graphicx,epsfig}

\usepackage{mathrsfs}
\usepackage{amsmath,amssymb}
\usepackage{epsfig}
\usepackage{relsize}
\usepackage{graphicx}

\usepackage{epsfig}
\usepackage{relsize}

\usepackage{mathrsfs}
\usepackage{amsmath,amssymb}
\usepackage{epsfig}

 \csname
@addtoreset\endcsname{equation}{section}

\def\bec{\begin{center}}
\def\ec{\end{center}}

\def\D{\Delta}

\def\del{\partial}

\def\tr{{\rm tr}}

 \def\det{{\rm det\,}}
\def\be{\begin{equation}}
\def\ee{\end{equation}}
\def\bea{\begin{eqnarray}}
\def\eea{\end{eqnarray}}
\def\ba{\begin{array}}
\def\ea{\end{array}}

\def\gc _{g_{_{\chi}}}

\newcommand{\comment}[1]{}

\author{Dmitri Bykov 
 \\ {\it School of Mathematics,
Trinity College, Dublin 2, Ireland} \\ {\it Steklov
Mathematical Institute, Moscow, Russia \;}\footnote{Emails:
dbykov@maths.tcd.ie, dbykov@mi.ras.ru}}

\abstract{We consider the $AdS_4 \times \mathbb{C}P^3$ IIA superstring sigma-model in the background of the "spinning string" classical solution, which possesses two Noether spins. In the limit when one of the spins is infinite there are massless excitations, which govern the infrared worldsheet properties of the model. We obtain a sigma-model of $\mathbb{C}P^3$ with fermions, which describes the dynamics of these massless modes.}

\title{The worldsheet low-energy limit \\of the $\mathrm{AdS_4 \times \mathbb{C}P^3}$ superstring}

\preprint{
          \smaller{\smaller{\smaller{TCDMATH 10-02}}}}

\dedicated{Dedicated to the anniversary of my Teacher \\Professor Andrey Alekseevich Slavnov}

\begin{document}
\section{Introduction}\label{intro}

Despite decades of investigation, the quantum mechanical behaviour of gauge theories is not fully understood. The main challenge is of course to shed light on the low-energy dynamics of four-dimensional QCD. Among the questions which still lack a thorough theoretical description are: confinement, chiral symmetry breaking, mass gap generation, as well as the precise relation of asymptotic freedom to the above mentioned properties. One of the tools which might eventually lead to an answer to these puzzles is the AdS/CFT correspondence \cite{m}. So far it has been convincingly formulated for a limited class of conformal field theories, among which the most important one is the Yang-Mills theory with $\mathcal{N}=4$ supersymmetry in four dimensions. Apart from the latter, other examples were put forward \cite{abjm,zarembonew}. It has been argued that the theory of \cite{abjm} might be useful for condensed matter studies, however, from our point of view the main reason for an intense study of this theory, to which we will contribute in this paper as well, is the desire to understand the range of applicability of the AdS/CFT correspondence in general. Gradually this might lead to an understanding of the AdS/CFT correspondence "from the first principles of gauge theory".

In the present paper we will be dealing solely with the string sigma-model, so we will not elaborate on the precise structure of the dual gauge theory. The interested reader can find the details (including the Lagrangian) in the original paper \cite{abjm}. Nevertheless, we would like to mention that the dual gauge theory is a three-dimensional conformal theory of Chern-Simons type with $\mathcal{N}=6$ supersymmetry. It was conjectured in \cite{abjm} to be dual to the $AdS_4 \times \mathbb{C}P^3$ type IIA superstring theory. The conformal group in the three-dimensional Minkowski space is $SO(2,3)$, and the R-symmetry group of the theory is $SU(4)$, which coincides with the isometry group of the string theory $AdS_4 \times \mathbb{C}P^3$ background. The number of supersymmetries is the same on both sides too, being equal to 24 (so these theories are not maximally supersymmetric, the maximum number of real supercharges being 32).

Similarly to the $AdS_5 \times S^5\;{\rm vs.}\;\mathcal{N}=4 \;{\rm SYM}$ case various integrability properties were found on both sides of the correspondence. Integrability of the two-loop scalar sector was found in \cite{z1} and refined in \cite{z2}. Subsequently an all-loop Bethe ansatz was proposed \cite{gv}, as well as the S-matrix \cite{an}. On the string theory side, the Green-Schwarz action was built and the Lax pair was constructed in \cite{af}. In this framework the so-called "off-shell" symmetry algebra was calculated, and in particular its central extension was shown to coincide with the one of the $AdS_5 \times S^5$ case \cite{Bykov}. The pp-wave limit for this background was first discussed in \cite{nt}, and the worldsheet near-pp-wave properties were examined in \cite{Sundin,h}. There also exists a pure spinor formulation of the $AdS_4 \times \mathbb{C}P^3$ superstring \cite{bonelli}.

It has been known for a long time \cite{gross} that the anomalous dimensions of operators of the form $ \tr(\Psi D_{+}^{S} \Psi)$ with $S\to \infty$ (that is, with large spin $S$) are of the form $\Delta = f(\lambda) \log(S)$, where $f(\lambda)$ is some function of the 't Hooft coupling constant $\lambda$, which in principle can be determined order by order in perturbation theory. In the context of the $\mathcal{N}=4$ supersymmetric Yang-Mills theory the spin $S$ enters the conformal $SO(2,4)$ part of the superconformal algebra. Hence, after the advent of the AdS/CFT correspondence it was realized by \cite{gkp} that the characteristic $\log{S}$ behaviour can be reproduced on the sigma model side of the correspondence by a particular classical solution of the worldsheet string theory in the $AdS$ background (which we will call the GKP solution in what follows). Indeed, provided one takes the correspondence for granted, their solution can even explain this behaviour.

If one quantizes the Green-Schwarz action of the superstring in the background of such a spinning string solution, one can find the spectrum of masses of the worldsheet particles (this problem was first solved in the $AdS_5 \times S^5$ setup in \cite{ft, ftt}). Some of them are massive, whereas others are massless. As emphasized by Alday and Maldacena \cite{am} for the case of $AdS_5 \times S^5$, there're certain quantities in the sigma model which receive the greatest contribution from the massless particles and their interactions. This is what we call "the worldsheet low-energy limit". Thus, it is an interesting question, what these massless particles are and what their interactions are in the $AdS_4 \times \mathbb{C}P^3$ case. The first part of the question was answered in \cite{aab,mr}, and here we will answer the second half. It turns out that there're 6 massless bosons and one Dirac fermion, their dynamics being described by the following Lagrangian:
\be\label{final}
\mathcal{L} = \,\eta^{\alpha\beta}\, \overline{\mathcal{D}_\alpha z^j} \, \mathcal{D}_\beta z^j  \,+ \,i 
\overline{\Psi} \gamma^\alpha\widehat{\mathcal{D}}_\alpha \Psi +{1\over 4} (\overline{\Psi} \gamma^{\alpha} \Psi)^2,
\ee
where index \(j\) runs from 1 to 4, $\mathcal{D}_\alpha= \partial_\alpha-i\,\mathcal{A}_\alpha$, $\widehat{\mathcal{D}}_\alpha= \partial_\alpha+2\, i\,\mathcal{A}_\alpha$ and $\mathcal{A}_\alpha$ is a $U(1)$ gauge field without a kinetic term --- it can be integrated out to provide the conventional Fubini-Study form of the action. Besides, in (\ref{final}) the $z^j$ fields are restricted to lie on the $S^7 \subset \mathbb{C}^4$:
\be
\sum\limits_{j=1}^4 |z^{j}|^2=1
\ee

The paper is organized as follows. In Section \ref{spinstring} we describe the spinning string classical solution of the sigma-model. In Section \ref{cosetsec} we describe the coset construction of the Green-Schwarz action, following \cite{af}. In Section \ref{AMlimit} we focus on the Alday-Maldacena limit in the familiar case of $AdS_5 \times S^5$. The discussion of Sections \ref{cosetsec} and \ref{AMlimit} emphasizes the reason, why the coset construction does not suffice for the consideration of an analogous limit in the present case. In Section \ref{GSaction} we elaborate on the construction of the full GS action with 32 fermions for the present case, following \cite{s1,s2}. In Section \ref{quantum} we explain various properties of the expansion around the spinning string solution. Finally, in Section \ref{low-energy} we find the low-energy limit of the expanded action, and find the sought for Lagrangian of the $\mathbb{C}P^3$ sigma model with fermions (\ref{final}). In the Appendix the interested reader will find details of the calculations.

\section{The spinning string}\label{spinstring}

The GKP solution \cite{gkp} describes a string moving in the $AdS_3$ subspace of the entire space. It is due to this reason that it is meaningful in the $AdS_4 \times \mathbb{C}P^3$ case as much as in the $AdS_5 \times S^5$ case. The solution is most easily described in the "global" coordinates, that is the coordinates which cover the whole of $AdS$ space. It is well-known that the $AdS$ space can be described as a hyperboloid embedded in the $\mathbb{R}^{2,D-1}$ space, namely (for the $AdS_4$ case) the surface
\be
-X_0^2 - X_1^2+X_2^2+X_3^2+X_4^2=-R^2
\ee
embedded into $\mathbb{R}^{2,3}$ with metric $ds^2=-dX_0^2-dX_1^2+dX_2^2+dX_3^2+dX_4^2$. The parameter $R$ is the "radius" of the $AdS$ space and describes its curvature. If one writes $X_0=\cosh(\rho) \cos(T),\;X_1=\cosh(\rho) \sin(T)$ and introduces spherical coordinates for $X_2, X_3, X_4$, the radius of the sphere being $\sinh(\rho)$, one obtains the global parametrization of the $AdS_4$ space. The metric then obtains the following form:
\be
(ds^2)_{AdS_4}\equiv G_{\mu\nu} dY^{\mu} dY^{\nu}=R^2 \left(-\cosh^2(\rho) dT^2+d\rho^2+\sinh^2(\rho) d\Omega_2 \right) .
\ee
To consider the $AdS_3$ space one simply needs to change $ d\Omega_2 \to d\phi^2$. In these coordinates the spinning string ansatz may be written in the following form:
\be
T = \kappa \tau,\; \phi = \omega \tau,\; \rho=\rho(\sigma)
\ee
If one denotes by $G_{ab} = G_{\mu\nu} \partial_a X^{\mu} \partial_b X^{\nu}$ the pull-back of the target-space metric, the Virasoro conditions are:
\be
G_{ab}-{1\over 2} \gamma_{ab} \gamma^{cd} G_{cd}=0.
\ee
As is customary, only two of them are independent. Let us impose the conformal gauge, then the Virasoro conditions are $G_{00}+G_{11}=0,\;G_{01}=0$. The latter is trivially satisfied by the ansatz, whereas the former produces an equation
\be \label{vir}
\rho'(\sigma)^2-\cosh^2(\rho) \,\kappa^2+\sinh^2(\rho) \,\omega^2=0.
\ee
One can check that the equations of motion for $T$ and $\phi$ are satisfied identically, whereas the one for $\rho(\sigma)$ coincides with the $\sigma$ derivative of (\ref{vir}).  The general solution $\rho(\sigma; \kappa, \omega)$ of (\ref{vir}) can be written in terms of elliptic functions. One should recall that the solution for the closed string is also subject to the periodicity condition \[\rho(\sigma+ 2 \pi;  \kappa, \omega)=\rho(\sigma;  \kappa, \omega), \] (which can be satisfied if one assumes that the string is folded) and this condition relates $\kappa$ to $\omega$ \cite{gkp,ft}. In the limit $\kappa \to \infty$, however, it turns out that $\omega = \kappa + ...$, so the solution simplifies drastically: 
\be \label{1spin}
T=\kappa \tau,\; \phi=\kappa \tau,\; \rho=\pm \kappa \sigma +\rho_0
\ee
The limit $\kappa \to \infty$ is called the long string limit, since, as one can see from the solution above, the extent of the string in $AdS$ space becomes infinitely large.

Solutions of the equations of motion can be classified according to the values of their Cartan charges, for example in this case one has\footnote{$g$ is the string tension, and it is related to the 't Hooft coupling constant $\lambda$ via $g\sim {\sqrt{\lambda}}$.}:
\bea
E = g \int\limits_{0}^{2\pi}\, \cosh^2(\rho)\,\dot{T}\,d\sigma=\kappa \,g\,\int\limits_{0}^{2\pi}\, \cosh^2(\rho)\,d\sigma, \;\\
S= g \int\limits_{0}^{2\pi}\, \sinh^2(\rho)\,\dot{\phi}\,d\sigma= \kappa \,g\,\int\limits_{0}^{2\pi}\, \sinh^2(\rho)\,d\sigma .
\eea
For every nontrivial solution the function $T(\tau,\sigma)$ is non-constant, since otherwise the Virasoro condition $G_{00}+G_{11}=0$ would require all target-space coordinates to be constant, thus virtually every solution possesses the $E$ charge. Due to this, one usually refers to a solution with a nonzero $S$ charge as a one-spin solution.

Clearly, the solution (\ref{1spin}) has one parameter $\kappa$, but it is convenient to use $S(\kappa)$ as a genuine parameter, since it has a more clear meaning. Then the target-space energy $E$ becomes a function of $S$, so let us calculate this function. First of all, $E-S=2\pi \kappa g$. Besides,
\be
S \underset{\kappa \to \infty}{\to} \frac{g}{2} \,e^{\pi \kappa}
\ee
It follows that
\be
\kappa \approx {1\over \pi} \log({S\over g}),
\ee
or in other words
\be
E-S\approx 2g \log({S\over g}),
\ee
so indeed the $\log(S)$ behaviour is reproduced. Besides, we immediately get a prediction that in the strong coupling limit $f(\lambda)\approx 2g$. What we are interested in are the quantum corrections to this function --- these are the $1/g$ corrections in the sigma model setup.

\subsection{The two-spin solution}\label{2spinstring}

The two-spin solution is a generalization of the spinning string solution described above, which possesses two nonzero Noether charges instead of one. One of these charges is the same $S$ charge, which originates from the motion in $AdS$ space, whereas the second one, called $J$, describes motion in the $\mathbb{C}P^3$ part of the space. This motion is in fact very simple and is just rotation around a big circle $S^1$ (which we will parametrize by $\varphi$) inside of $\mathbb{C}P^3$. Thus, the ansatz looks as follows:
\be \label{2spin}
T=\kappa \tau,\; \phi=\omega_1 \tau,\; \varphi = \omega_2 \tau , \; \rho=\rho(\sigma)
\ee
Then the nontrivial Virasoro condition gives, instead of (\ref{vir}):
\be
\rho'(\sigma)^2-\cosh^2(\rho) \,\kappa^2+\sinh^2(\rho) \,\omega_1^2 +\omega_2^2=0.
\ee
Once again, the solution simplifies tremendously in the limit $\kappa \to \infty$, when $\omega_1 \approx \kappa$, so $|\rho'(\sigma)|=\sqrt{\kappa^2-\omega_2^2}$.

The charge $J$ has the following value:
\be
J=g\int\limits_{0}^{2\pi}\,\dot{\varphi}\,d\sigma=2\pi g \omega_2 .
\ee

It will be convenient for further use to introduce a parameter
\be\label{upar}
u=\frac{2 \omega_2}{\kappa}= \frac{J}{g \,\log(S)}
\ee

\section{The coset $OSP(6|4) \left/ \right. U(3)\times SO(1,3)$}\label{cosetsec}

The target superspace, in which the movement of the string occurs, can in fact be described as a coset $OSP(6|4) \left/ \right. U(3)\times SO(1,3)$. The bosonic part of this coset is $SO(6)\left/ \right. U(3) \times SP(4) \left/ \right. SO(1,3)$, which in fact is the desired space $AdS_4 \times \mathbb{C}P^3$. Besides the bosonic part of the superspace, the coset includes 24 real fermions. For a simple description of a matrix realization of the corresponding groups the reader is referred to \cite{af}, and a review of similar ideas for the case of the $AdS_5 \times S^5$ string can be found in \cite{review}.

The coset construction of the Green-Schwarz superstring action is rather simple. The key to this simplicity lies in the fact that the $OSP(6|4)$ superalgebra possesses a $Z_4$ group of automorphisms (this is a cyclic group, generated by an element which we call $\Omega$). Then, suppose we take a representative $g(x,\theta)$ of the coset ($\theta$ here represent the fermions, and $x$ the worldsheet coordinates), and build the left-invariant current
\be
J=-g^{-1}\,dg(x,\theta).
\ee
Then one can determine the 4 components of this current, which lie in the eigenspaces of the $\Omega$ transformation. Let us denote them by $J^{(k)},\,k=0,1,2,3$, and their characteristic property is $\Omega J^{(k)} \Omega^{-1}= i^{k}\,J^{(k)}$. Then the action invariant under the $\Omega$ automorphism (or, equivalently, under the $Z_4$ group of automorphisms) is built uniquely in the following way:
\be\label{cosetaction}
S= \frac{g}{2}\,\int\,d\sigma\,d\tau\,\left(\gamma^{\alpha\beta}\, {\rm Str}(J^{(2)}_{\alpha}\,J^{(2)}_{\beta})+\varkappa \,\epsilon^{\alpha\beta}\,{\rm Str}(J^{(1)}_{\alpha} \,J^{(3)}_{\beta})\right).
\ee
When $\varkappa =\pm 1$ this action possesses an important gauge symmetry, called kappa-symmetry, which is a fermionic gauge symmetry in the sense that the gauge parameters are anticommuting.

\subsection{Expansion around the two-spin solution}\label{expansion}

One can impose the conformal gauge and expand the coset action around the two-spin solution (\ref{2spin}) and determine the spectrum of the fluctuation fields. This was done in \cite{aab,mr}, where the following result was obtained: when $\omega_2=0$, all the bosonic fields from $ \mathbb{C}P^3$ are massless, two of the $AdS$ fields have masses $m^2=2\kappa^2$ and $m^2=4\kappa^2$, and the remaining two $AdS$ fields are massless, but their contribution is supposed to cancel against the ghost contribution (in other words, they do not contribute to the cohomology of the BRST operator). As for the fermions, for a generic value of $\omega_2$ the determinant of the fermionic quadratic form in the action was found to be:\bea \nonumber \label{det}
\mathscr{D}&=&2^8\omega_2^{16}\big[(2k_0-\omega_2)^2-4(k_1^2+\varkappa^2)\big]^2\big[(2k_0+\omega_2)^2-4(k_1^2+\varkappa^2)\big]^2
\times\\
&&~~~~~~~~~~~~~~~~~~~~~~~~~~~~\times
\big[k_0^4-k_0^2(2k_1^2+\varkappa^2)+k_1^2(k_1^2-\omega_2^2+\varkappa^2)\big]^2\,
. \eea
It follows from this expression that there are the following fermionic excitations in the model (counting given in terms of complex Weyl fermions, and we're using the parameter $u$ introduced in (\ref{upar})):
\begin{itemize}
  \item 2 fermions with frequency $\frac{1}{4} \,u \,\kappa+\sqrt{n^{2}+\kappa^{2}}\, $
  \item 2 fermions with frequency $-\frac{1}{4} \,u \,\kappa+\sqrt{n^{2}+\kappa^{2}}\, $
  \item 2 fermions with frequency $\sqrt{n^{2}+{ \kappa^{2} \over 2}(1+ \sqrt{1+u^{2}n^{2}})}$
  \item 2 fermions with frequency $\sqrt{n^{2}+{ \kappa^{2} \over 2}(1- \sqrt{1+u^{2}n^{2}})}$
\end{itemize}

The spectrum in the background of the {\it spinning string} solution is obtained when $u\to 0$ (or $\omega_2 \to 0$), and we see that the spectrum becomes relativistic in this limit: there are 6 massive and 2 massless complex fermions, i.e. 12 massive and 4 massless real ones. However, one can then see from (\ref{det}) that the quantization in the background of the {\it spinning string} solution, that is in the limit $u\to 0$, is no longer well-defined, since the determinant is zero. This corresponds to the fact that, as noted for the first time in \cite{af}, the kappa symmetry transformations degenerate at the quadratic level, if the background configuration describes motion purely in the $AdS$ space. We would like to stress that this is a peculiarity of the coset construction for the $AdS_4 \times \mathbb{C}P^3$ superstring. One can rephrase the statement above by saying that the quadratic part of the action will contain terms of the form $u^2\,\psi \partial_{\pm} \psi$ ($\partial_\pm$ being the light-cone derivatives), which will vanish when $u=0$. One could be tempted to rescale the fermions $\psi \to \frac{1}{u}\psi$, however it turns out that in this case $1/u$ factors would appear in front of the interaction terms. Thus, in this way we would merely shift the problem to a different place, and this only confirms that the difficulty is intrinsic to the coset formulation.

In fact, as we have explained, the most interesting for us are the massless worldsheet modes, i.e. the ones which become massless in the limit $u\to 0$. We have checked that it is precisely these massless fermions $\psi$, which suffer from the problem descibed above, namely the quadratic action of these fermions becomes degenerate in the limit $u\to 0$. Hence, it seems to us that everything points out to the fact that the coset formulation is not well suited to this background. Therefore we will have to resort to the full Green-Schwarz action with 32 fermions, which was built in \cite{s1,s2} (see \cite{s3} for another application of that construction), in order to solve our problem. 

The problem we have encountered is, in fact, connected with a singular gauge choice, and similar issues arise also in conventional gauge theories, as described in Appendix B.

\section{The worldsheet low-energy limit of the $AdS_5\times S^5$ superstring}\label{AMlimit}
 
Here we describe the limit introduced in \cite{am}\footnote{I am grateful to Kostya Zarembo for clarifying to me some details of this limit.}. In the previous section, after quantization of the superstring worldsheet action in the background of the spinning string solution we obtained a spectrum of masses of the worldsheet particles present in the theory. This spectrum becomes particularly simple in the limit $u\to 0$, namely it becomes relativistic, and some of the particles are massless. It is this limit which was emphasized in the paper \cite{am}, their main idea being as follows. One considers the partition function (or, equivalently, free energy) of the worldsheet theory in the background of the two-spin solution, regarding $u$ as a variable parameter, which in fact is the chemical potential for one of the global charges of the model. The limit of small $u$, that is when $\omega_2 \ll \kappa$, is effectively the low-energy limit, and one expects the massless particles to give dominant contributions to the free energy, and therefore in this sense the massless particles "decouple". In the $AdS_5 \times S^5$ case the only massless particles are the 5 bosons coming from the $S^5$ part of the background, so their dynamics is described by the $SO(6)$ sigma model, which decouples from the rest of the theory in this limit. From the exact solution of the $SO(6)$ model it is known that, in fact, its spectrum consists of 6 massive particles rather than 5 massless ones, however the mass gap cannot be seen in perturbation theory ($m \sim e^{-a g}$), which is still applicable as long as $\omega_2 \gg m$.

To be more precise, the leading contributions to the free energy in the limit $u\to 0$ look as follows:
\be
E\propto u^2\,\left( g + \sum\limits_{n=1}\,{1\over g^{n-1}}\,(a_{n}\,(\log u)^n +... ) \right)\ee
One can see that the power of the logarithm grows with the order of ${1\over g}$, which is a common feature of perturbation theory. The main quantitative claim of \cite{am} was that in the $AdS_5 \times S^5$ case the numbers $a_n$ can be determined from the pure $SO(6)$ sigma model, and it was confirmed in \cite{rots} up to two loops.

In the $AdS_5 \times S^5$ case  the only massless particles are bosons from $S^5$, and their interactions are determined by the $SO(6)$ symmetry, so the model is uniquely defined by these properties. In the $AdS_4 \times \mathbb{C}P^3$ case there are additional massless fermions, as we discussed above, and the amount of symmetry is not enough to determine their interactions uniquely. Because of that a genuine calculation is necessary, and it is carried out below.

\section{The full IIA superstring action in the $AdS_4 \times \mathbb{C}P^3$ background}\label{GSaction}

As explained in the previous sections, for the analysis of the action, expanded around the spinning string solution, one needs to use the complete type IIA Green-Schwarz superstring action (that is, with 32 real fermions), rather than the reduced coset formulation (in the $AdS_5\times S^5$ case the coset action is the complete Green-Schwarz action --- it contains 32 fermions --- and it was first built in \cite{mt}). The construction of such action is no easy task and for generic supergravity backgrounds it has not been carried out. However, as we will explain, the $AdS_4 \times \mathbb{C}P^3$ case under consideration is special, since it can be obtained by a dimensional reduction of the $AdS_4 \times S^7$ solution of the eleven-dimensional supergravity equations of motion. One might wonder, why this would simplify anything. However, from the work \cite{bst} it is known that there exists a three-dimensional world-volume action of a membrane, coupled to an arbitrary eleven-dimensional supergravity background (usually this membrane is called the M2 brane). This action is quite similar to the superstring action in many ways, for example in the sense that it, too, possesses a local fermionic symmetry. It was argued in \cite{dhs}, that if one compactifies a target-space coordinate and a world-sheet coordinate of this action simultaneously, then one obtains the Green-Schwarz type IIA superstring action. Thus, the remaining question is whether it is easy or not to build the M2 brane worldvolume action. The answer to this question depends on the chosen supergravity background, but in our case the task simplifies, since the $AdS_4 \times S^7$ background can be described by a coset $OSP(8|4) \left/ \right. SO(7) \times SO(1,3)$. Indeed, its bosonic part is $SP(4)\left/ \right. SO(1,3) \times SO(8)\left/ \right. SO(7)  \approx AdS_4 \times S^7$. This strategy was pursued in the papers \cite{s1,s2}, where as a result the sought for type IIA Green-Schwarz action was built. Here we elaborate on this construction in a, hopefully, transparent way.

\subsection{The M2 brane action and the coset $OSP(8|4) \left/ \right. SO(7) \times SO(1,3)$}\label{M2brane}

In this section we describe the construction of the M2 brane three-dimensional worldvolume action for the case of the $AdS_4 \times S^7$ supergravity background. In what follows we will denote the membrane worldvolume coordinates as $\sigma, \tau$ and $y$. We remind the reader that the fields in the eleven-dimensional supergravity are the graviton $g_{\mu\nu}$, the gravitino $\psi^{\alpha}$ and the three-form potential $\mathcal{H}^{(3)}$ (the field strength being the four-form $\mathcal{F}^{(4)}=d\mathcal{H}^{3}$). In our construction we heavily exploit the $OSP(8|4) \left/ \right. SO(7) \times SO(1,3)$ coset structure of the space. The $OSP(8|4)$ supergroup is very similar to the $OSP(6|4)$ supergroup, and its matrix realization is described in the Appendix \ref{osp}. To construct the explicit matrix realization of the coset one also needs to choose the embeddings of "the denominator" $SO(7) \times SO(1,3)$ into this supergroup. For instance, $SO(7)$ can be embedded into $SO(8)$ in different nonequivalent ways (here we mean, that the embeddings are in general not related by a similarity transformation). We will elaborate more on this in the Appendix \ref{so6embed}, however for the moment let us give a clear 

{\it Example}. One can embed $SO(7)\subset SO(8)$ diagonally (that is, as a $7\times7$ matrix inside of a $8\times8$ matrix), let us denote this embedding $h_1 : SO(7) \to SO(8)$. There's another, "spinorial", embedding. Indeed, let $\gamma^{\mu},\,\mu=1...7$ denote the real skew-symmetric seven-dimensional gamma-matrices ($\{\gamma^{\mu},\gamma^{\nu}\}=-2\delta^{\mu\nu}$), which have dimensionality 8. Then the commutators $\gamma^{\mu\nu}={1\over 2}[\gamma^{\mu},\,\gamma^{\nu}]$ generate the $so(7)$ algebra inside of $so(8)$. We denote the corresponding group embedding as $h_2$. Now, $h_1$ and $h_2$ cannot be related by a similarity transformation. Indeed, let $z$ be a fixed element of the $so(7)$ algebra, such that $h_2(z)$ is nondegenerate, for instance we can take $z=h_2^{-1}(\gamma^{12})$. If there were a similarity transformation relating the two embeddings, it would imply that $h_1(z)=\omega\,h_2(z)\,\omega^{-1}$ for some $\omega \in SO(8)$. However, since $h_1$ is the diagonal embedding, $\det(h_1(z))=0$, whereas $\det(h_2(z))\neq 0$, which leads to a contradiction. $\rhd$

We stress that the correct embedding for our purposes is the "spinorial" one, that is the $so(7)$ algebra is generated by $\gamma^{\mu\nu}$.

The M2 brane action can be built from the bosonic and fermionic vielbeins, denoted by $E^{A}=\{E^{\bar{a}},\, E^{a}\}$ and $E^\alpha$ respectively, which in turn can be obtained in a simple way from the $OSP(8|4) \left/ \right. SO(7) \times SO(1,3)$ coset. Here the indices $\{\bar{a}=0...3,\,a=1...7\}$ refer to the $AdS_4$ and $S^7$ spaces respectively, and the index $\alpha=1...32$ numbers the fermionic directions. The number 32 comes from the fact that we're dealing with a single Majorana spinor in eleven dimensions, which has 32 real components. If considered from the point of view of the representation of the ten-dimensional Lorentz group, it splits into two Majorana-Weyl spinors --- one left-handed and one right-handed. Let $g$ be a representative of the coset. Then one can build the left-invariant current $J=-g^{-1}\,dg$ and find the vielbein and connection components of this current:
\be
J=-g^{-1}\,dg=E^{a} T_a + E^{\alpha} Q_{\alpha} + A^{ab} \Omega_{ab},
\ee
where $\Omega_{ab}$ are elements of the stabilizer (denominator of the coset), $Q_{\alpha}$ furnish the fermionic basis of the $osp(8|4)$ algebra, and $T_a$ are the complementary bosonic directions, namely the directions tangent to the manifold. The current is flat by construction, that is its curvature vanishes:
\be\label{maurer}
dJ-J\wedge J=0.
\ee

The action, as in the superstring case, consists of two terms, which can be loosely called 'the metric part' and the 'Wess-Zumino part'. The action takes the following form:
\be\label{action11d}
S=\frac{g}{2\pi}\,\int\,d\sigma\,d\tau\,dy\,\left(\eta_{ab}\,E^{a}E^{b}+\varkappa \,\epsilon^{abc} \,\partial_{a}X^{m}\,\partial_{b}X^{n}\,\partial_{c}X^{p}\,\mathcal{H}_{mnp} \right),
\ee
the first term being the metric part and the second term --- the Wess-Zumino part\footnote{By the coordinates $X^m$ we mean both bosonic and fermionic ones.}. The second term is the pull-back to the worldvolume of a three-form $\mathcal{H}=\mathcal{H}_{mnp}\,dX^{m}\wedge dX^{n}\wedge dX^{p}$. Recall that this three-form can be found from its field-strength four-form $\mathcal{F}$, and the latter can be expressed in terms of the vielbeins in the following way\footnote{ $\widehat{\lambda}$ is a constant. Barred indices refer to the AdS space, that is they run from 0 to 3.}:
\be\label{WZmembrane}
\mathcal{F}={1\over 8}\epsilon_{\bar{a}\bar{b}\bar{c}\bar{d}} E^{\bar{a}} \wedge E^{\bar{b}} \wedge E^{\bar{c}} \wedge E^{\bar{d}}+\widehat{\lambda}\, E^{\alpha} \wedge [\mathbf{\Gamma}_{A},\mathbf{\Gamma}_{B}]_{\alpha}^{\beta}\,E_{\beta}\wedge E^{A}\wedge E^{B}.
\ee
The first term here is the volume element of the $AdS$ space, whereas the second term is intrinsically fermionic and manifestly $Spin(1,10)$ invariant. Note that the fermionic indices ($\alpha,\beta$) are raised and lowered using the eleven-dimensional charge conjugation matrix $C_{11}$. Using the Maurer-Cartan equation (\ref{maurer}) one can check that the form $\mathcal{F}$ is closed for a suitable value of the constant $\widehat{\lambda}$. A closed form is locally exact, so one can find its potential $\mathcal{H}$ by a standard procedure. We will follow this route in the next sections to build the part of the Green-Schwarz action that we need.

\subsection{The Hopf fibration $S^7 \to \mathbb{C}P^3$ and the dimensional reduction}\label{hopfbundle}

So far we have been dealing with the $AdS_4 \times S^7$ solution of the eleven-dimensional supergravity. However, our ultimate goal is to arrive at the $AdS_4 \times \mathbb{C}P^3$ solution of the IIA supergravity in ten dimensions. This is achieved through a compactification, based on the Hopf fibration, so we recall what the latter looks like. There are different variants of the Hopf fibration, most of them originating from the tautological fiber bundle $\mathbb{C}^{n+1}\to \mathbb{C}P^n$ (the one which arises naturally from the definition of the projective space). The most common version of the Hopf fibration arises when one restricts the total space $ \mathbb{C}^{n+1}$ to the unit sphere $\sum\limits_{i=1}^{n+1} |z_{i}|^2=1$, then we get $\pi: S^{2n+1}\to \mathbb{C}P^n$. The tautological bundle is a line bundle (its fiber is $\mathbb{C}$), so after imposing the absolute value restriction the $\pi$ fibration has the circle $S^1$ as a fiber. The most well-known case of the Hopf fiber bundle is explained in the following

{\it Example.} It is the case $n=1$, i.e. $\pi: S^{3} \to  \mathbb{C}P^{1}\approx S^2$. If we write the sphere $S^3$ as 
\be \label{3sphere}
|z_1|^2+|z_2|^2=1,
\ee
then the map $\pi$ can be written out explicitly as $\pi(z_1,z_2)=\frac{z_2}{z_1}\equiv Z$, and $Z$ should be regarded as a stereographic coordinate on the sphere $S^2$. To see what is going on more clearly, let us solve (\ref{3sphere}) in the following way: $z_1 = \frac{y_1}{\sqrt{|y_1|^2+|y_2|^2}},\,z_2 = \frac{y_2}{\sqrt{|y_1|^2+|y_2|^2}}$. The metric on the sphere is the metric, induced in flat space $ds^2=|dz_1|^2+|dz_2|^2$ by the embedding (\ref{3sphere}). Thus it is clear that it will be invariant under arbitrary rescalings $y_i \to \lambda y_i,\,\lambda \in \mathbb{R}$. In other words, when regarded as a metric on the 4-space parametrized by $y_1, y_2$, it is degenerate. This means, of course, that we can fix a "gauge", however we do not want to do it for the moment. We can rewrite the metric in the following way:
\be\label{hopfmetric1}
(ds^2)_{S^3}= \frac{dy_{i} d\bar{y}_{i}}{\rho^2}-\frac{|dy_{i} \bar{y}_{i}|^2}{\rho^4}+\left(i\frac{dy_{i} \bar{y}_i-y_i d\bar{y}_i}{2 \rho^2}\right)^2,\,\textrm{where}\,\,\rho^2=|y_i|^2 .
\ee
 The first two terms constitute precisely the $\mathbb{C}P^1$ metric in homogeneous coordinates. Let us now choose the following inhomogeneous coordinates:
 \be \label{hopfcoord}
 y_{1}=e^{i\varphi},\;y_{2}= e^{i\varphi}\, Z,
 \ee
 then the metric takes the canonical form:
 \be\label{hopfmetric2}
 (ds^2)_{S^3}= \frac{dZ d\bar{Z}}{(1+Z\bar{Z})^2}+(d\varphi-A)^2, \;\textrm{where}\;\,\, A=i\frac{dZ \,\bar{Z}-Z\,d\bar{Z}}{1+Z\bar{Z}}.
 \ee
 Note that from (\ref{hopfcoord}) it follows that, in accordance with our discussion, $Z= \frac{y_2}{y_1}=\frac{z_2}{z_1}$. $\rhd$
 
Although the space which interests us in this paper is $ \mathbb{C}P^3$ rather than $ \mathbb{C}P^1$, it can be considered in a similar way to the example, and as a result one obtains a metric on $S^7$ in the form, which exhibits the Hopf bundle in a clear way. This discussion shows, that the dimensional reduction of the metric can be achieved by dropping the $(d-A)^2$ term in the metric, or getting rid of the corresponding einbein. Note also that, when fermions are turned on, $A$ is interpreted as the R-R one-form $\mathcal{A}^{(1)}$ of the IIA background. As for the four-form $\mathcal{F}$, it can be written in the following way:
\be
\mathcal{F}=\mathcal{K}^{(3)}\wedge d\varphi+\mathcal{G}^{(4)},
\ee
where $\mathcal{K}^{(3)}$ and $\mathcal{G}^{(4)}$ do not depend on the fiber coordinate $\varphi$. Then the dimensional reduction (after setting $\varphi = y$, $\;y$ being the third worldvolume coordinate  \cite{dhs}) boils down to leaving the $\mathcal{K}^{(3)}$ piece of the $\mathcal{F}$ form: $\textrm{dim.red.}(\mathcal{F})=\mathcal{K}^{(3)}$. One can see that, since $d\mathcal{F}=0$, it follows that $d\mathcal{K}^{(3)}=d\mathcal{G}^{(4)}=0$. In fact, it follows from our discussion that $\mathcal{K}^{(3)}$ is the field-strength of the (locally defined) NS-NS two-form potential $\mathcal{B}^{(2)}$, which constitutes the Wess-Zumino term of the string action: $\mathcal{K}^{(3)}=d\mathcal{B}^{(2)}$. We note in passing that $\mathcal{D}^{(4)}=\mathcal{G}^{(4)}-\mathcal{A}^{(1)}\wedge \mathcal{K}^{(3)}$ is the R-R 4-form field-strength of the IIA background (we will not need it later). In this way we have determined all the ingredients of the ten-dimensional IIA supergravity solution from the eleven-dimensional supergravity solution. 
 
Since the construction we will carry out heavily relies on the $OSP(8|4)$ coset, we need to parametrize the coset in such a way which would exhibit the Hopf fiber bundle, similarly to {\ref{hopfmetric2}}. Compared to the purely bosonic case, the main difference is that various reductions can be made, which preserve different amounts of supersymmetry. The correct one, i.e. the one we're looking for, should provide for a $OSP(6|4)$ symmetry group of the {\it reduced} background. The reductions differ by the embedding of $OSP(6|4)\subset OSP(8|4)$, so in the Appendix \ref{so6embed} we explain, following \cite{s1,s2}, that the correct one is the standard diagonal embedding. 

\section{Quantum corrections to the spinning string state}\label{quantum}

\subsection{The decompactification}\label{decompact}

The Green-Schwarz action in its standard form does not allow for a simple quantization of the theory. This is due to the fact that the action does not contain a term quadratic in the fermions with no bosons, but rather terms, in which fermions are coupled to bosons. The quantization is possible in the background of a classical solution, if the classical solution makes the quadratic term of the fermions nondegenerate. As we discussed above, this requirement is fulfilled for the two-spin solution even in the $OSP(6|4)$ coset formulation, however in the limit when the $J$ charge vanishes this is no longer the case. It is precisely to overcome this difficulty that we needed to invoke the full Green-Schwarz action with 32 fermions. Once we have built the full action, we may quantize in the background of the spinning string (one spin) solution (\ref{1spin}) in the limit $\kappa \to \infty$. It will be convenient however to make a change of the worldsheet coordinates:
\be
\sigma' = \kappa \sigma,\;\;\tau' = \kappa \tau .
\ee
The important observation is that after such redefinition the only parameter which is changed is the length of the worldsheet circle, or the string length: $L=2\pi \kappa$. In other words, $\kappa$ enters only the integration limits in the action, but not anywhere else. Since (\ref{1spin}) is a solution for $\kappa \to \infty$, in this limit we achieve the decompactification of the worldsheet.

\subsection{The coset element for the spinning string background.}\label{spinstringcoset}

We take the following $AdS_3$ coset element:
\be
g_{AdS_3}= e^{\frac{i}{2} \,t \,\Gamma^{0}-\frac{1}{2} \,\phi \,\Gamma^{1} \Gamma^{2} } \,e^{-\frac{1}{2}\,\rho \,(i \Gamma^{2}-\Gamma^{0} \Gamma^{1} ) }
\ee
When we insert the spinning string solution (\ref{1spin}), it simplifies as follows:
\be\label{gspin}
g_{spin}= e^{\frac{1}{2}\, ( i\Gamma^{0}- \Gamma^{1} \Gamma^{2}) \,\kappa \tau } \,e^{-\frac{1}{2}\,\rho(\sigma) \,(i \Gamma^{2}-\Gamma^{0} \Gamma^{1} ) }
\ee
The important property of this parametrization is that $[i \Gamma^{0}- \Gamma^{1} \Gamma^{2}, i \Gamma^{2}-\Gamma^{0} \Gamma^{1}]=0$, and it is precisely thanks to this fact that the current $g^{-1} dg$ only includes the derivative $\rho'=\pm \kappa$, but not $\rho$ (and hence not $\sigma$) itself. The latter fact is the reason why it is possible to determine the spectrum of quadratic fluctuations at all, and it is what makes the theory in this background tractable.

\subsection{The expansion}\label{spinstringexp}

The full coset element in the spinning string background looks as follows:
\be \label{fullcoset}
g=g_{spin}\,g_{AdS}\,g_{\mathbb{C}P}\,g_{\theta}\,e^{\varphi \mathcal{\epsilon}}\,g_{v},
\ee
where $g_{AdS}$ describes fluctuations of the $AdS$ fields, $g_{\mathbb{C}P}$ is the $\mathbb{C}P^3$ coset element, $\varphi$ is the fiber coordinate, $g_{\theta}$ and $g_{v}$ are the fermionic coset elements. In particular, $g_{\theta}$ only includes the 24 fermions which were present in the $OSP(6|4)$ coset (see (\ref{theta}) for $\theta$), whereas $g_v=e^{v_{\lambda} Q^{\lambda}}$ contains the 8 additional fermions (see (\ref{vferm}) for $v$). It will be convenient to take the coset elements $g_{AdS}$ and $g_{\mathbb{C}P}$ in the following form:
\bea\label{adsfluct}
&g_{AdS}= e^{{(z_+-z_-)\over 2}(i \Gamma_0 - \Gamma_1 \Gamma_2)}\,e^{{(z_+ + z_-)\over 2}(i \Gamma_2 - \Gamma_0 \Gamma_1)}  \,\frac{1+{i\over 2}(z_1 \Gamma_1+z_2 \Gamma_3)}{\sqrt{1-{1\over 4}(z_1^2+z_2^2)}}&\\
&g_{\mathbb{C}P}=1+\frac{W+\overline{W}}{\sqrt{1+|w_i|^2}}+\frac{\sqrt{1+|w_i|^2}-1}{|w_j|^2 \sqrt{1+|w_i|^2}}\left(W\overline{W}+\overline{W} W\right),&
\eea
where $W=w_i \mathcal{T}_i,\,\overline{W}=\overline{w}_i \overline{\mathcal{T}}_i,$ and one can find definitions of $\mathcal{T}_i, \overline{\mathcal{T}}_i$ in Appendix \ref{projspace}. Note that the two exponents in (\ref{adsfluct}) commute with each other, as well as with $g_{spin}$ of (\ref{gspin}), which is the reason why we have chosen the parametrization in this way (as a result, the current $J=-g^{-1} dg$ will depend on $z_\pm$ only through their derivatives).

We need to plug the coset element into the expression for the action of the 11D theory (\ref{action11d}), fix the kappa symmetry gauge and to determine, which of the fermions are massive and which are massless. One might recall however that, as determined before, the spectrum consists of 12 massive and just four massless fermions. Thus, let us use the symmetry properties of the theory as a shortcut to the result. Indeed, if one multiplies the coset element (\ref{fullcoset}) by an element $\omega \in SU(3)$ from the left, then we get the following:
\be
\omega g = g_{spin}\,g_{AdS}\,(\omega\,g_{\mathbb{C}P}\,\omega^{-1})\,(\omega\,g_{\theta}\,\omega^{-1})\,e^{\varphi \mathcal{\epsilon}}\,g_{v}\,\omega,
\ee
where $\omega$ at the very right can be dropped, as it belongs to the stabilizer of the coset (the $SU(3)$ generators are the $L_i$'s of (\ref{l1}-\ref{l3}), and, since they're linear combinations of $\gamma^{\mu\nu}$, it follows that $SU(3)\subset SO(7)$). Clearly $\omega\,g_{\mathbb{C}P}\,\omega^{-1}$ simply transforms the bosonic fields of the model, whereas $\omega\,g_{\theta}\,\omega^{-1}$ transforms the fermionic $\theta$ fields. We see that the transformation does not affect the $v$-fermions. The masses of the theory are determined by the quadratic part of the Lagrangian, in which the bosonic and fermionic fields clearly decouple. In fact, we might forget about the bosonic fields for the moment, since we already know their spectrum. Let us remind the reader that the matrix of $\theta$ fermions can be written as follows:
\begin{equation}\label{theta}
\theta = \begin{bmatrix}
\theta^{1}_{1}&\theta^{1}_{2}&\theta^{1}_{3}&\theta^{1}_{4}&\theta^{1}_{5}&\theta^{1}_{6}&0&0\\
\theta^{2}_{1}&\theta^{2}_{2}&\theta^{2}_{3}&\theta^{2}_{4}&\theta^{2}_{5}&\theta^{2}_{6}&0&0\\
\star&\star&\star&\star&\star&\star&0&0\\
\star&\star&\star&\star&\star&\star&0&0
\end{bmatrix},
\end{equation}
where the stars stand for complex conjugated fermions.

The action $\omega\,g_{\theta}\,\omega^{-1}$ means, that the $\theta^{1}$ and $\theta^{2}$ fermions (6 complex in each line) furnish $3+\bar{3}$ representations of $SU(3)$ each. The important point is that none of the $\theta$ fermions are invariant under the $SU(3)$, whereas all of the $v$ fermions are. For the quadratic part of the action this means that the $\theta$ and $v$ fermions decouple. Now, a $3$ or $\bar{3}$ representation of $SU(3)$ involves 6 real fermions, so if the $SU(3)$ symmetry can be preserved by a choice of the kappa symmetry gauge, this means that every $6$ $\theta$-fermions entering a single multiplet have the same mass. A suitable kappa symmetry gauge is the one, which sets the second and third lines of the $\theta$ and $v$ matrices to zero (from the Appendix \ref{low-energyapp} it is clear that such gauge is indeed admissible). This condition can be summarized as follows:
\be\label{kappagauge}
\frac{I-i \Gamma_{0}\Gamma_{1}\Gamma_2}{2}\,\theta=\frac{I-i \Gamma_{0}\Gamma_{1}\Gamma_2}{2}\,v=0.
\ee
We know that there are just 4 massless fermions, so the $SU(3)$ multiplets are too big for that, and the $\theta$ fermions are destined to be {\it massive}. Thus, we're left with the 8 $v$-fermions. The residual kappa-symmetry should allow to eliminate 4 of them, and the remaining 4 ones should be {\it massless}. We have checked by a direct calculation of the quadratic part of the action that this is indeed true.

\section{The worldsheet low-energy limit}\label{low-energy}

In the previous section it was explained that, before the kappa symmetry gauge is imposed, the $\theta$- and v-fermions decouple in the quadratic action. In fact, one can check that in the leading (linear) order the kappa symmetry transformations do not mix $\theta$'s and v's. This is the reason, why it is possible to choose the kappa-gauge in such a way, that $\theta$'s and v's remain decoupled in the gauge-fixed quadratic action. By now we know that the $\theta$ fermions (meaning the ones that remain after the kappa gauge is imposed) are all massive, so we can safely set them to zero. Then we are left with an action depending solely on the bosonic fields and the v-fermions, and there will be a residual kappa symmetry of rank 4 acting on these fields. One could certainly gauge-fix this residual symmetry as well from the beginning, however it is useful to check that the resulting low-energy action which we will obtain is independent of the kappa-symmetry gauge choice. For this reason we will prefer not to fix the gauge for the v-part of the kappa-symmetry.

First of all, let us find out, how exactly this v-part of the kappa-symmetry looks like in the leading order. For this purpose we write out the piece, which depends on v's, of the quadratic part of the string action\footnote{The notations for the fermions are explained in Appendix \ref{matrices}, 'v-fermions'.}:
\be\label{quadVpart}
\mathcal{L}_v^{(2)}= i (\bar{\chi}-\bar{\xi})\del_+ (\chi-\xi) - i(\bar{\psi}-\bar{\eta})\del_- (\psi-\eta)
\ee
It follows that the kappa-transformations have a very simple form:
\be\label{kappav}
\delta \chi = \delta \xi = \epsilon_1,\;\;\delta \psi = \delta \eta = \epsilon_2,
\ee
$\epsilon_{1,2}$ being two Weyl spinors. Suppose we now want to determine the low-energy limit of the worldsheet theory. What terms should be left from the full Green-Schwarz action in order to achieve this? Clearly, one should get rid of the massive fields. Besides, one should also drop certain interaction vertices of the massless fields (the ones that are suppressed by powers of some mass), since they, too, may be regarded as effectively reflecting the presence of massive modes (in this way, for example, the Fermi four-fermion interaction appears in the Standard Model after the $W$ and $Z$ boson fields have been integrated out --- this interaction is suppressed by the masses of the bosons and should be dropped in the strict low-energy limit). In other words, all terms in the low-energy Lagrangian should have dimension not greater than two. In 2D the bosons $w$ have canonical dimension $0$, the fermions have dimension $1/2$ and the derivative, clearly, has dimension $1$. In this way, for instance, terms of the form $\psi_{1}\,\psi_2\,\chi_1\,\del_+ \chi_1$ and $\psi_1 \psi_2 \del_+ w\,\del_- \bar{w}$ have dimension $3$, terms $\psi_{1}\,\del_-\psi_2\,\chi_1\,\del_+ \chi_2$ have dimension 4, so such terms should be dropped in the low-energy limit. Most of the terms which should be preserved, have dimension two: these are, for instance, $\del_+ \bar{w} \del_- w,\; \bar{\chi}\del_+ \chi,\;\bar{w} \del_- w\, \bar{\psi} \psi,\; \bar{\psi} \psi \bar{\chi} \chi$ etc. However, a very important fact which should not be overlooked, is that there are terms of dimensions $1$ and $0$ in the Lagrangian as well: the terms of dimension $1$ are $z \, v^2$, $z$ being the massive $AdS$ fields, and the terms of dimension $0$ are the mass terms $z^2$. Hence, there's a very important qualification to dropping the massive fields: they can be dropped everywhere, except for these terms. What is the meaning of these terms or, in other words, what is the reason for their being present in the final Lagrangian? It turns out that these terms are precisely what is needed to maintain the kappa-symmetry of the low-energy Lagrangian, that is to say they provide the independence of the low-energy Lagrangian of the chosen kappa-gauge. One can see this in the following way. In the lowest order the kappa-symmetry transformations look similar to the case of the flat target space: 
\be\label{kappatr}
\delta v =\epsilon,\;\;\delta z = \epsilon v
\ee
We will not bother writing any indices or projectors here, since this schematic exposition is sufficient to convey the general idea. From (\ref{kappatr}) it follows that
\be\label{kappatr1}
\delta (v^4) \propto \epsilon v^3,\;\;\delta (z^2) \propto \epsilon v z,\;\;\delta(z\,v^2) \propto \epsilon v^3 + \epsilon v  z .
\ee
Our claim is that the three terms $z^2$, $z\,v^2$ and $v^4$ enter the Lagrangian in a kappa-invariant combination, which is allowed by the rules (\ref{kappatr1}). There's another way to make the same point, namely, let us integrate out the massive fields $z$. What we obtain as a result are terms of the form $v^4$ ($v$ here can mean any of the 8 different $v$-fields, so there are many ways to build $v^4$ terms). Then, our statement can be reformulated by saying that all the $v^4$ terms should combine into one term
\be
(\bar{\chi}-\bar{\xi}) (\chi-\xi) (\bar{\psi}-\bar{\eta}) (\psi-\eta)
\ee
which is the only quartic combination of the $v$-fields, invariant under the transformations (\ref{kappav}.) We have checked that this is indeed the case, and the reader can find the details of the calculation in the Appendix \ref{low-energyapp}.

Once we have checked the independence of the resulting Lagrangian of the choice of the kappa-symmetry gauge, we may impose the one which we find most convenient (\ref{kappagauge}) to obtain the following Lagrangian (see Appendix \ref{matrices} for fermion field notations):
\be\label{lagr2}
\mathcal{L} = {1\over 2} \, G_{\bar{a}b}(z,\bar{z}) \,(\partial_+ z_{a} \partial_-\bar{z}_b+\partial_- z_{a} \partial_+ \bar{z}_b)+ i (\bar{\chi} \mathcal{D}_+ \chi -\bar{\psi} \mathcal{D}_- \psi) +3\, \bar{\chi}\chi\bar{\psi}\psi ,
\ee
where  $\chi=\psi_1+i\chi_1,\;\psi=\psi_2+i\chi_2$ and the  "covariant" derivatives are:
\be
\mathcal{D}_{\pm}=\del_{\pm}+ \,\frac{\bar{z_i} \del_\pm z_i - z_i \del_\pm \bar{z}_i}{|z_{j}|^2}
\ee
Besides, $G_{\bar{a}b}$ is the Hermitian ($G_{\bar{a}b}=G_{\bar{b}a}^{\ast}$) Fubini-Study metric on $\mathbb{C}P^3$:
\be
G_{\bar{a}b}=\frac{\delta_{\bar{a}b}}{|z_{j}|^2}-\frac{\bar{z}_{a}\,z_{b}}{|z_{j}|^4}
\ee

Upon the introduction of a Dirac field \( \Psi =  \left(
\psi \atop \chi \right) \) the Lagrangian we have obtained may be cast into the form announced in the Introduction \footnote{The $\gamma$-matrices here are the 2D gamma-matrices, for instance, in our notations $\gamma^0=-i \sigma^2, \gamma^1=\sigma_1, \gamma^5 = \sigma_3$ and $\mathcal{D}_\pm = \mathcal{D}_0 \pm \mathcal{D}_1$. They should not be confused with the 7D $\gamma$ matrices from the Appendix and other parts of the text. The conjugation is defined as $\overline{\Psi}=\Psi^\dag \gamma^1$.}:
\be\label{lagr3}
\mathcal{L} = \, \eta^{\alpha\beta}\, \overline{\mathcal{D}_\alpha z^j} \, \mathcal{D}_\beta z^j  \,+ \,i 
\overline{\Psi} \gamma^\alpha\widehat{\mathcal{D}}_\alpha \Psi +{1\over 4} (\overline{\Psi} \gamma^{\alpha} \Psi)^2,
\ee
where index \(j\) runs from 1 to 4, $\mathcal{D}_\alpha= \partial_\alpha-i\,\mathcal{A}_\alpha$, $\widehat{\mathcal{D}}_\alpha= \partial_\alpha+2\, i\,\mathcal{A}_\alpha$ and $\mathcal{A}_\alpha$ is a $U(1)$ gauge field without a kinetic term --- it can be integrated out to provide the conventional Fubini-Study form of the action. Besides, in (\ref{lagr3}) the $z^j$ fields are restricted to lie on the $S^7 \subset \mathbb{C}^4$:
\be
\sum\limits_{j=1}^4 |z^{j}|^2=1
\ee
Note that the Lagrangian in (\ref{lagr3}) is invariant under global $U(1)\times U(1)$ transformations of the fermions, that is $\Psi \to e^{i \alpha} \Psi$ and $\Psi \to e^{i  \alpha \gamma^5} \Psi$.

As it had been expected, we have obtained a fermionic extension of the nonlinear sigma-model with target space $\mathbb{C}P^3$.

\section{Open problems}\label{concl}

We have obtained the Lagrangian (\ref{lagr3}) for a model describing the infrared limit of the worldsheet theory of the $AdS_4\times \mathbb{C}P^3$ superstring, quantized in a certain background. This Lagrangian can be used to calculate certain quantitative characteristics of the model, as discussed in Section \ref{AMlimit}. The latter could be compared with the Bethe ansatz predictions (as it has been done in \cite{g1} for the $AdS_5 \times S^5$ case).

Apart from the string theory applications, the model we have obtained might be interesting in its own right. In the past a great deal of effort was devoted to the understanding of various $\mathbb{C}P^3$ models, with and without supersymmetry \cite{div,wit,zum}, since it was hoped they could give some insight into the infrared dynamics of QCD. These models are asymptotically free, however it is only the bosonic model that exhibits confinement, whereas models with fermions usually describe liberated $U(N)$ solitons. In most, if not all, cases the quantum S-matrix of such solitons can be computed exactly. It is an interesting question, whether the model we have obtained is integrable as well. One could also wonder, whether there is any fundamental explanation for the $6:2$ ratio of the bosonic vs. fermionic degrees of freedom. There are many questions which remain to be answered. We plan to address them in a subsequent publication.

\acknowledgments{I am grateful to Sergey Frolov for suggesting that I work on the problem of the low-energy limit and for numerous useful and illuminating discussions in the course of work. I am also indebted to Dmitry Sorokin for a very patient email correspondence explaining his work, as well as to Kostya Zarembo for a useful discussion of several aspects of the present paper. I also want to thank Sergey Frolov, Dmitry Sorokin, Per Sundin and Kostya Zarembo for carefully reading the manuscript and contributing to its improvement by valuable comments. Besides, I am glad to congratulate my Teacher Professor A.A.Slavnov on the occasion of his anniversary, to thank him for his constant support and encouragement, and to wish him many more years of prolific work and interesting life. My work was supported by the Irish Research Council for Science, Engineering and Technology, in part by grants RFBR 08-01-00281-a, 09-01-12150-ofi\_m and in part by grant for the Support of Leading Scientific Schools of Russia NSh-8265.2010.1.}

\appendix

\section{The projective space}\label{projspace}

$\mathbb{C}P^3$ is usually defined as the space of ratios $(z_1 : z_2 : z_3 : z_4)$ in $\mathbb{C}^4/\{0\}\;$ \footnote{In this section we follow the exposition of \cite{Bykov}.}. However, there are other ways to look at this space. One of them is to view $\mathbb{C}P^3$ as the space of orthogonal complex structures in $\mathbb{R}^6$. Indeed, $U(3)\subset O(6)$ is the subgroup preserving a given complex structure, which we denote $K_6$ and, following \cite{af}, choose in the form $K_6 = I_3 \otimes i \sigma_2$ ($I_3$ is the $3\times 3$ identity matrix). Then the Lie subalgebra $u(3)\subset o(6)$ is described by $6\times 6$ matrices, commuting with $K_6$. In other words, as vector spaces, $o(6)=u(3)\oplus V_\perp$.

The quotient vector space $W$, which describes the tangent space $T_x \mathcal{M}$ (tangent spaces are isomorphic for all $x$, since $\mathcal{M}$ is a manifold), is described by skew-symmetric matrices (elements of $o(6)$, that is) which anticommute with the complex structure. Indeed, we notice that for any $\omega \in O(6)$ the adjoint action $\omega K_6 \omega^{-1}$ is again a complex structure. For $\omega$ sufficiently close to unity $\omega=1+\epsilon + ...$, thus, $(K_6+[\epsilon, K_6])^2 +O(\epsilon^2) = -I_6$. Linear order in $\epsilon$ gives $\{K_6, [\epsilon, K_6]\}=0$. Define a map $f: o(6) \to o(6)$ by $f(a)=[a,K_6]$. Since ${\rm Ker}(f)={\rm u}(3)$, $W$ is isomorphic to ${\rm Im}(f)$. 
One can also check that if $g(b)\equiv\{K_6,b\}=0$, then $b\in{\rm Im}(f)=W$. \footnote{In fact, this choice of representatives in the quotient space becomes canonical once we adopt the Killing scalar product (since $f$ is skew-symmetric with respect to this scalar product $\tr(a,f(c))=-\tr(f(a),c)$). Indeed, for $a\in u(3)$ and $b\in {\rm Im}(f)$ we have $\tr (ab) = \tr(a[c,K_6])=\tr(a c K_6-a K_6 c)=0$, since $[a,K_6]=0$). This justifies the use of the symbol $V_\perp$ for $W$.} Let us note in passing that all of the above can be summarized by the following exact sequence of vector space homomorphisms ($i$ being inclusion):
\begin{equation}\label{exactseq}
0\to u(3)\overset{i}{\to} o(6) \overset{f}{\to} o(6) \overset{g}{\to} \mathbb{R}^{N},
\end{equation}
$\mathbb{R}^{N}$ being the vector space of symmetric matrices.

 It is easy to construct a basis in this linear space explicitly. Denoting by $J_1, J_2, J_3$ the three generators of $O(3)$ in the vector $3$ representation
 \begin{equation}\small
J_{1}=
\begin{pmatrix}
0&1&0 \\ -1&0&0 \\ 0&0&0
\end{pmatrix},\qquad
J_{2}=
\begin{pmatrix}
0&0&1 \\ 0&0&0 \\ -1&0&0
\end{pmatrix},\qquad
J_{3}=
\begin{pmatrix}
0&0&0 \\ 0&0&1 \\ 0&-1&0
\end{pmatrix},
\end{equation}
we get: 
\begin{equation}
V_\perp = \textrm{Span} \{ J_i \otimes \sigma_1 ; J_i \otimes \sigma_3 \}
\end{equation}
To make contact with the notations of \cite{af} we will write out the $T_i$ generators used in their paper in terms of the basis introduced above:
\begin{equation}
T_{1,3,5}=  J_{1,2,3} \otimes \sigma_3,\; T_{2,4,6}= J_{1,2,3} \otimes \sigma_1 .
\end{equation}
The main property which these generators exhibit and which will be important for us is the following:
\begin{equation}
\{T_1, T_2\}=\{T_3, T_4\}=\{T_5, T_6\}=0.
\end{equation}
In the body of the paper we used the following complex combinations:
\begin{equation}
\mathcal{T}_1 = \frac{1}{2} (T_1 - i T_2),\;\;\mathcal{T}_2=\frac{1}{2} (T_3 - i T_4),\;\;\mathcal{T}_3=\frac{1}{2} (T_5 - i T_6).
\end{equation}
$\bar{\mathcal{T}}_1$, $\bar{\mathcal{T}}_2$ and $\bar{\mathcal{T}}_3$ denote the conjugate combinations.

\section{The particle spectrum in different gauges: an example}\label{partspectrum}

The problem we described in Section \ref{cosetsec} is related to a singular gauge choice for a part of the kappa-symmetry.

To clarify the situation we present an example from the hard core of gauge theory, where a similar phenomenon occurs. Namely, we will consider the Abelian $U(1)$ Higgs model with the standard Lagrangian:
\be\label{Higgs}
\mathcal{L}_{{\rm Higgs}}=-{1\over 4} F_{\mu\nu}^2+\overline{D}^{\mu}\phi^{\ast}\,D_{\mu}\phi-{\widehat{g}\over 4} (\phi^{\ast} \phi-v^2)^2
\ee
In the above clearly $\phi$ is the Higgs field, the covariant derivative is $D_{\mu}\phi=(\partial_{\mu}+i \,g\,A_{\mu})\,\phi$ and the gauge transformations are \be\nonumber\phi \to e^{i\,g\,\alpha}\,\phi,\;A_{\mu}\to A_{\mu}-\partial_{\mu} \alpha\ee It is possible to choose the so-called "unitary" gauge, which corresponds to setting $\phi$ to be real. In this gauge the Lagrangian (\ref{Higgs}) obtains the following form:
\be\label{HiggsUnitary}
\mathcal{L}_{{\rm Higgs}}= -{1\over 4} F_{\mu\nu}^2+(\partial_{\mu}\phi)^2+g^2\,A^2 \phi^2-{\widehat{g}\over 4} (\phi^2-v^2)^2
\ee
Now, as long as $v\neq 0$ in order to stabilize the potential $V={\widehat{g}\over 4} (\phi^2-v^2)^2 $ one usually makes a shift $\phi=v+\varphi$, which, among other things, produces the following quadratic form:
\be\label{HiggsQuad}
\mathcal{L}_{{\rm Higgs}}^{(2)}= -{1\over 4} F_{\mu\nu}^2+(\partial_{\mu}\varphi)^2+g^2 v^2 A^2-\widehat{g}\,v^2\, \varphi^2
\ee
In particular, for $v\neq 0$ the quadratic form above is nondegenerate, its zeros describing the spectrum of the theory: 3 particles of mass $m_1 = gv$ (which come from the gauge field $A_\mu$) and 1 particle of mass $m_2 = \widehat{g}^{1/2}v$ (which comes from the scalar $\varphi$). One should thus expect that in the limit $v\to 0$ we would obtain 4 massless particles. However, in practice this limit is rather subtle, and this is due to the fact that the quadratic Lagrangian (\ref{HiggsQuad}) becomes gauge-invariant in the limit $v\to 0$, despite the fact that a gauge has already been chosen. This is of course an artifact of the combination of gauge choice and the perturbation expansion, since the gauge invariance is broken, as it should be, by the interaction terms that we have dropped in (\ref{HiggsQuad}). The same statement can be reformulated, if one looks at the propagator $D_{\mu\nu}$ of the gauge field
\be
D_{\mu\nu}(k)=\frac{1}{k^2-g^2 v^2}\,\left(\eta_{\mu\nu}-\frac{k_\mu k_\nu}{g^2 v^2}\right),
\ee
which clearly is singular of order $\sim 1/v^2$ when $v\to 0$. Thus, the situation is similar to the case we are considering in this paper, since, as explained in Section \ref{cosetsec}, the propagators of some of the fermions behave as $1/u^2$ in the $u^2\to 0$ limit (equivalently, their quadratic Lagrangian is proportional to $u^2$).

What is the solution to this problem? According to the general logic explained above, we need to find a more suitable gauge, so that the quadratic part of the Lagrangian is nondegenerate even in the $v\to 0$ limit. There are many gauges at our disposal, for instance the Feynman gauge, in which the quadratic Lagrangian looks as follows (when $v=0$):
\be
\mathcal{L}_{{\rm Higgs}}^{(2)}= -{1\over 2} (\partial_{\mu}A_{\nu})^2+|\partial_{\mu}\phi|^2+\bar{c}\,\Box \,c,
\ee
$c,\,\bar{c}$ being the Faddeev-Popov ghosts. In particular the cohomology of the BRST $Q$ operator in the $A_\mu$ sector consists of two states, which describe the two polarizations of the massless vector field. Apart from them, we also have the two massless scalar fields $\phi,\,\bar{\phi}$, thus the spectrum of the model indeed consists of 4 massless particles. Notice, however, that in this case two of these particles come from $A_\mu$ and two from $\phi$, whereas in the case of the unitary gauge we had three particles from $A_\mu$ and one from $\phi$. The different ways of splitting the sectrum is of course a natural consequence of gauge invariance of the model. The important point is that both approaches give the same spectrum in the limit $v\to 0$, however only one of them is applicable for perturbative calculations.

In using the quite involved construction described in the paper we have in mind the simple idea of choosing a proper gauge for our model.

\section{The $osp(8|4)$ superalgebra}\label{osp}

This section of the Appendix provides a matrix realization of the $osp(8|4)$ superalgebra. The discussion here is in many ways parallel to the one of \cite{af}, since the $osp(6|4)$ algebra described there is very similar to the one of our interest.

Generators of the $osp(8|4)$ can be thought of as $4|8 \times 4|8$ supermatrices:
\be
A=\left(\begin{array}{cc}
X&\;\vartheta\\
\eta &\; Y
\end{array}
\right),
\ee
where $X$ and $Y$ are bosonic matrices of dimensions $4\times4$ and $8\times 8$ respectively. The matrix $Y$ belongs to $so(8)$ and, as such, is real and antisymmetric:
\be
Y^\ast=Y,\;\;Y^{\rm T}=-Y
\ee
The matrix $X$ belongs to $sp(4)$ and can be characterized by the following properties:
\be
X^{\ast}= i \Gamma_2 \,X\, (i \Gamma_2)^{-1},\;\;X^{\rm T} = -C_4 \,X\, C_4^{-1}
\ee
As for the fermions, $\eta$ is related to $\vartheta$ via
\be
\eta = -\vartheta^{\rm T} C_4
\ee
and the reality property reads
\be
\vartheta^\ast=i\Gamma_2 \vartheta
\ee

\section{Embeddings $SO(6) \hookrightarrow SO(8)$}\label{so6embed}

The purpose of this section of the Appendix is to prove that the standard diagonal embedding of $OSP(6|4)\subset OSP(8|4)$ is the one relevant for our purposes.

A Lie superalgebra can be decomposed into its bosonic and fermionic components in a standard way: $\mathcal{L}=\mathcal{L}_0 +\mathcal{L}_1$. Then, $\mathcal{L}_0$ is represented on $\mathcal{L}_1$, since $[\mathcal{L}_0 , \mathcal{L}_1]\subset \mathcal{L}_1$. Thus, a natural question arises which representations arise in this way. The answer to this question (among others) was given by Kac \cite{kac}. For the case of the $osp(8|4)$ superalgebra the corresponding fermionic module is $sp_4 \otimes so_8$, where $sp_4$ and $so_8$ are the standard (defining) representations of the corresponding algebras. In other words, for practical purposes one can consider the representation in terms of $8|4 \times 8|4$ supermatrices:
\begin{equation}
M=
\left(
\begin{array}{ccc}
  A & B   \\
  C  & D  
\end{array}
\right),
\end{equation}
where $A$ is a standard representation of $sp(4)$ and $D$ is the standard representation of $so(8)$ \footnote{It is well-known that $Spin(8)$ has 3 different irreps of dimension 8, related by the so-called triality, so in this case the theorem of Kac rules out two of them, which are the chiral and anti-chiral spinorial ones. }, whereas $B$ and $C$ are the fermionic algebra elements, subject to natural reality properties (besides, $C$ is conjugate to $B$, in the sense that it is uniquely determined by the latter). In fact, representations of both of these bosonic algebras can be conveniently described in terms of gamma-matrices. Let us denote by $\Gamma_{\mu}, \mu=0,1,2,3$ the $D=4$ gamma-matrices, and by $\gamma_{\alpha}, \alpha=1... 6$ (and also $\gamma_7$ on slightly separate grounds) the $D=6$ (respectively $D=7$) gamma-matrices. They satisfy the following defining conditions:
\begin{eqnarray}
\{\Gamma_{\mu},\Gamma_{\nu}\}=2 \eta^{(4)}_{\mu\nu},\;\;\eta^{(4)}={\rm diag} (+,-,-,-)\\
\{\gamma_{\alpha},\gamma_{\beta}\}=2 \eta^{(6)}_{\alpha \beta},\;\;\eta^{(6)}={\rm diag} (-,-,-,-,-,-)
\end{eqnarray}
In these signatures the $\Gamma$ matrices may be chosen to be imaginary, and the $\gamma$ matrices may be chosen to be real (this is the choice of Majorana bases for both Clifford algebras\footnote{Note however that throughout the paper we used a different representation of the $\Gamma$ matrices, see Appendix \ref{matrices}.}). The $sp(4)$ algebra is then generated by \cite{af} ${1\over 2}[\Gamma_{\alpha},\Gamma_{\beta}],\;\; i \Gamma_{\alpha}$, so all the generators are real. The $so(8)$ algebra is generated by standard $E_{ij}$ matrices (with $1$ on the $ij$-place and $-1$ on the $ji$ place), so (not surprisingly) it is real too. In this setup the fermions should be chosen real as well.

There are different ways to represent $SO(6)$ on the 8-dimensional vector space of the $SO(8)$ vector representation. One of them is the standard diagonal embedding, which can be continued to the embedding $OSP(6|4) \subset OSP(8|4)$ in a simple way:
\begin{eqnarray}
G_{OSP(6|4)}=\left(
\begin{array}{cc | c}
  A & B\; & \\
  C  & D\; & \\
  \hline
  & & \;1_2
\end{array}
\right),
\end{eqnarray}
The other representation is a faithful\footnote{We remind the reader that a representation $r$ of a group $\mathcal{G}$ on a vector space $V$, that is a homomorphism $r: \mathcal{G}\to \mathrm{GL}(V)$, is called faithful, if $r$ is injective.} $Spin(6)$ representation (and, as such, not a representation of $SO(6)$). It may be constructed in the following way. Let $x_{1}...x_{8}$ be the 8 coordinates in the vector space, on which $SO(8)$ is represented. We can form complex combinations $X^{\pm}_{1}=x_{1}\pm i x_{2}$, etc. Then, those $SO(8)$ transformations which correspond to analytic (linear) maps of $X^{+}_{1,2,3,4}$ form an $SU(4)$.  In this way $SU(4)=Spin(6)$ is represented irreducibly on the 8-dimensional real vector space (if one considered the vector space over $\mathbb{C}$, the representation of $SU(4)$ would split as $4+\bar{4}$). The above definition is equivalent to the following: one needs to choose those matrices from $SO(8)$ which commute with a given complex structure in $\mathbb{R}^8$. For definitiveness we choose the simplest complex structure, which in our conventions is given by $\gamma_7$. Clearly, such matrices are ${1\over 2}[\gamma_{\alpha},\gamma_{\beta}],\;\;\alpha,\beta=1...6$ (as well as $\gamma_7$ itself, which thus extends $SU(4)$ to $U(4)$). The latter can be split in two groups: the ones, which lie in $SU(3)\subset SU(4)$ (we call them $L_i$'s), and the ones which lie in the compliment (we call them $\mathbb{T}_i$ and $U$). These are the following:
\begin{eqnarray}\label{l1}
&L_1 = \gamma_{26}+\gamma_{15},\;\;L_2 = \gamma_{35} + \gamma_{46}\;\;L_3 = \gamma_{14} - \gamma_{23}& \\ \label{l2}
&L_4 = \gamma_{16} - \gamma_{25},\;\;L_5 = \gamma_{36} - \gamma_{45},\;\; L_6 = \gamma_{13} + \gamma_{24}&\\ \label{l3}
&L_7 = \gamma_{12} - \gamma_{34},\;\;L_8 = \gamma_{12} + \gamma_{34} - 2 \gamma_{56},\;\;U=\gamma_{12} + \gamma_{34} + \gamma_{56}.&\\
&\mathbb{T}_1 =1/2 ( \gamma_{26}-\gamma_{15}),\;\; \mathbb{T}_2 = 1/2 ( \gamma_{35} - \gamma_{46}),\;\;\mathbb{T}_3 = 1/2 (\gamma_{14} + \gamma_{23})&\\
&\mathbb{T}_4=1/2 (\gamma_{16} + \gamma_{25}),\;\; \mathbb{T}_5 = 1/2 (\gamma_{36} + \gamma_{45}),\;\; \mathbb{T}_6 = 1/2 (\gamma_{13} - \gamma_{24})&\end{eqnarray}
In the above $U$ is an element of $SU(4)$ which commutes with $SU(3)$.

In fact, the $SU(3)$ group, generated by $L$'s is the same, as the $SU(3)$ subgroup of the diagonal $SO(6)$ embedding. In the notations of the paper \cite{s1}, the projector $P_6 = {\rm diag} (1,1,1,1,1,1,0,0)$ leaves $L_i$ invariant $P_6 L_i P_6 = L_i$, and annihilates $\mathbb{T}_i$: $P_6 \mathbb{T}_i P_6 =0$.

In order to get the diagonal embedding in this way, one needs to project the $\gamma$-matrices (exactly as described in the appendix to paper \cite{s1}): $T_{i}= P_6 \gamma_i P_6 ,\;i=1...6$. We have chosen the Majorana gamma-matrices $\gamma$ in such a way, that these projections give precisely the $T$-matrices, defined in the paper \cite{af} (this is a confirmation that the embedding has been chosen correctly).

Once we know what the correct embedding is, we can proceed to define the Hopf fiber bundle. As described in \cite{s1}, in order to do this we need to represent the sphere $S^7$ as a coset $SU(4)\times U(1)/SU(3)\times U'(1)$, where the 'gauge group' $U'(1)$ is generated by the element $U$, whereas translation along the fiber $S^1$ is generated by $\gamma_7$. Note that one may write
\be\label{alpha}
\gamma_7=K_6 + \mathcal{\epsilon},
\ee
where $\mathcal{\epsilon}$ is defined in Appendix \ref{matrices} and
\be\label{beta} U=K_6-3 \mathcal{\epsilon} , \ee so the new space is indeed produced by a twisting of the original $U(1)$  gauge group generated by $K_6$ with the new direction $\phi$, that appears as the angle of $SO(2)$ (generated by $\mathcal{\epsilon}$) in $SO(6)\times SO(2) \subset SO(8)$, both subgroups embedded diagonally.

As explained in Appendix \ref{dimredapp}, dimensional reduction corresponds to dropping the einbein $e^{7}=d\phi-\mathcal{A}$, which describes the fiber. In terms of the coset, it implies an additional gauging of a $U(1)$ subgroup, generated by the fiber translations, that is by $\gamma_7$. On the other hand, as it follows from the formulas (\ref{alpha}, \ref{beta}) above, gauging both $U$ and $\gamma_7$ is the same as gauging $K_6$ and $\mathcal{\epsilon}$, and thus we return to the $\mathbb{C}P^3$ space, as we should. $\rhd$

\par Thus, the $OSP(8|4)/SO(7)\times SO(1,3)$ coset element can be chosen as follows:
\be\label{coset}
g= g_{OSP(6|4)}\,e^{\varphi \mathcal{\epsilon}}\,e^{v_{\lambda} \, Q^{\lambda}}.
\ee
Here $g_{OSP(6|4)}$ is the $OSP(6|4)$ coset element, which can be taken, for instance, from \cite{af}, and schematically it looks as follows: $g_{OSP(6|4)}=g_{bosons} \, h_{\,24 fermions}$. 
$v_\lambda$ are the additional 8 fermions absent in the $OSP(6|4)$ coset. The matrix $v_\lambda Q^{\lambda}$ can be found in the Appendix \ref{matrices}.

\section{Dimensional reduction in detail}\label{dimredapp}

This section of the Appendix is dedicated to the explanation of how the Kaluza-Klein reduction is performed in our setup. In part \ref{invariance} we prove that the reduction preserves the $OSP(6|4)$ subgroup of the $OSP(8|4)$ isometry group of the $AdS_4 \times S^7$ background.

\subsection{Metric term}\label{metricterm}

In this section we will follow the line of reasoning adopted in \cite{dhs}.

Suppose we have an 11D metric which can depend on the fermions as well as on the bosons. This metric is subject to an important qualification --- it has a linearly realized U(1) isometry, which we will take to be the shift $z\to z + a$, $a$ being an arbitrary constant. Then the metric can be written in the following way:
\be\label{metric}
ds^2 = g_{ab}dx^a dx^b + B_a dx^a dz + C dz^2
\ee
Assume that the three membrane coordinates are $\sigma, \tau, y$. Let us set $z=y$, then the pullback of the metric written above to the membrane worldvolume can be written as follows:
\be
\widehat{G}_{\alpha\beta} dx^\alpha dx^\beta = \left( g_{ab}\partial_\alpha x^a \partial_\beta x^b + B_a \partial_\alpha x^a \partial_\beta z + C \partial_\alpha z \partial_\beta z \right)  dx^\alpha dx^\beta
\ee
In the above formula the indices $\alpha, \beta$ run from 1 to 3. Now, when $\alpha, \beta = 1,2$ we have
\be
\widehat{G}_{\alpha\beta}= g_{ab}\partial_\alpha x^a \partial_\beta x^b
\ee
When $\beta=3$,
\be
\widehat{G}_{\alpha 3}={1\over 2} B_a \partial_\alpha x^a
\ee
Finally, if $\alpha=\beta=3$, we get
\be
\widehat{G}_{33}=C
\ee
If we calculate the determinant of $\widehat{G}_{\alpha\beta}$, it will of course be some function of $g_{ab}, B_a, C$. Since $\det(\widehat{G})\neq \det(g)$, $g$ cannot be regarded as the pullback of the correct 10D string metric. However, the following Kaluza-Klein construction cures this drawback. Indeed, the correct pullback $h_{ab}$ is intoduced by the following decomposition of $\widehat{G}$ (here $i=1,2$):
\bea\label{red}
\widehat{G}=\Phi^{-2/3}
\left(
\begin{array}{ccc}
h_{ij}+\Phi^2 A_i A_j  &  \Phi^2 A_i   \\
\Phi^2 A_j  & \Phi^2  
\end{array}
\right)
\eea
The point of this decomposition is that now for {\it any} $\Phi, A_i$ the following holds true:
\be
\det(\widehat{G})=\det(h)
\ee
Thus, the Nambu-Goto actions of the membrane and the string coincide up to a factor of the radius of the fiber, which we denote by $r$:
\be
\int d\sigma d\tau dy \sqrt{\det{\widehat{G}}} = r \int d\sigma d\tau \sqrt{\det{h}}.
\ee
One can read off the following from (\ref{red}):
\bea
&\Phi^{-2/3} \Phi^2 = \Phi^{4/3} = \widehat{G}_{33} & \\& \Phi^{-2/3} \Phi^2 A_i = \Phi^{4/3} A_i = \widehat{G}_{i3} & \\ & \Phi^{-2/3} (h_{ij}+\Phi^2 A_i A_j)= \widehat{G}_{ij}&
\eea
We need to express $\Phi^{4/3} , A_i$ and, ultimately, $h_{ij}$ from these expressions:
\bea
&\Phi^{4/3} = \widehat{G}_{33}&\\
&A_i = \frac{\widehat{G}_{3i}}{\widehat{G}_{33}}&\\
&h_{ij}= 
\sqrt{\widehat{G}_{33}}\,(\widehat{G}_{ij}-\frac{\widehat{G}_{3i} \widehat{G}_{3j}}{\widehat{G}_{33}})&
\eea
The last line of this equation gives us the sought for pullback (to the 2D string worldsheet) of the 10D metric. Its determinant is equal to the determinant of $\widehat{G}$, as explained above. It is a general answer, independent of the representation of the vielbeins (in this argumentation the vielbeins are irrelevant, since from the very beginning we're dealing with the metric, and of course one can always choose the vielbeins in a pretty form to satisfy the given metric).

The procedure we have just explained is equivalent to writing the original metric (\ref{metric}) in the form
\be\label{can}
ds^2 = \widetilde{g}_{ab}dx^a dx^b + C(dz-\widehat{B}_a dx^a)^2
\ee
then dropping the last term and multiplying the first one by an appropriate factor of $C$, such that the determinant is unchanged. 

\subsection{Wess-Zumino term}\label{wztermapp}

Here we explain the technical side of the Wess-Zumino term construction. It was announced in (\ref{WZmembrane}) that the Wess-Zumino term for the M2 brane action looks as follows:
\be\label{WZmembrane2}
\mathcal{F}={1\over 8}\epsilon_{\bar{a}\bar{b}\bar{c}\bar{d}} E^{\bar{a}} \wedge E^{\bar{b}} \wedge E^{\bar{c}} \wedge E^{\bar{d}}+\widehat{\lambda}\, E^{\alpha} \wedge [\mathbf{\Gamma}_{A},\mathbf{\Gamma}_{B}]_{\alpha}^{\beta}\,E_{\beta}\wedge E^{A}\wedge E^{B}.
\ee
The bosonic vielbeins in the above formula are normalized in a canonical way, namely so that the $AdS_4 \times S^7$ metric is written in the form $ds^2 = \eta_{11}^{AB} E_{A}E_{B}$.

The fermionic vielbeins $E^{\alpha}$ which enter this formula are the fermionic components of the supercurrent
\be
J=-g^{-1}\,dg=E^{a} T_a + E^{\alpha} Q_{\alpha} + A^{ab} \Omega_{ab},
\ee
We will not write out the matrix form of the supergenerators \(Q_\alpha \), since this is to a large extent irrelevant for our problem, but will rather deal with components of the matrix $\mathcal{Q}=E^{\alpha} Q_{\alpha}$. It is a supermatrix with zero bosonic components, so it has "off-diagonal" form. We will call its top right block $\Theta$ (its bottom left block $\widehat{\Theta}$ is related to it via  $\widehat{\Theta}=-\Theta^{{\rm T}}\,C_4$) and write its matrix components as $\Theta_{\bar{a}a}$, where $\bar{a}=0...3$ is the row number and $a=1...8$ is the column number.

After this preparational work we may write the term $E^{\alpha} \wedge [\mathbf{\Gamma}_{A},\mathbf{\Gamma}_{B}]_{\alpha}^{\beta}\,E_{\beta}$ as follows:
\[
E^{\alpha} \wedge [\mathbf{\Gamma}_{A},\mathbf{\Gamma}_{B}]_{\alpha}^{\beta}\,E_{\beta} = \Theta^{\bar{a}a} \wedge [\mathbf{\Gamma}_{A},\mathbf{\Gamma}_{B}]_{\bar{a}a}^{\bar{b}b} \,\Theta_{\bar{b}b} = \Theta_{\bar{a}a} (C_{11})^{\bar{a}a,\bar{b}b} \wedge [\mathbf{\Gamma}_{A},\mathbf{\Gamma}_{B}]_{\bar{b}b}^{\bar{c}c} \,\Theta_{\bar{c}c}
\]
The double-index notation is very convenient for our choice of the $\mathbf{\Gamma}$-matrices (see Appendix \ref{matrices}), since all of them have the form $A_4\otimes B_8$, and clearly $(A_4\otimes B_8)_{\bar{a}a}^{\bar{b}b}\equiv (A_4)_{\bar{a}}^{\bar{b}}\, (B_8)_a^b$. The $C_{11}$ matrix raises/lowers the indices, but the interpretation of its own indices is still the same: $(C_{11})^{\bar{a}a,\bar{b}b}= -(C_{11})_{\bar{a}a,\bar{b}b}=(C_4 \Gamma_5)^{\bar{a}\bar{b}} \delta^{ab}$ (the minus sign is due to the fact that $C_{11}^2=-1$). 

The requirement of the closedness of the four-form in (\ref{WZmembrane2}) fixes the value of $\widehat{\lambda}$:\be \widehat{\lambda}= {i\over 4}\ee.

\subsection{$OSP(6 \mid 4)$ invariance of the 10D theory}\label{invariance}

We mentioned above that the form (\ref{coset}) of the coset provides for the $OSP(6|4)$ invariance of the ten-dimensional theory. 
That is what we're going to prove here. For simplicity we will consider just the $\mathbb{C}P^3$ part of the problem, or the $SO(6)$ symmetry group. Let $\Omega\in SO(6)\subset SO(8)$ act on $g(w,\theta,v)$ ($w$ and $\theta,v$ are the bosonic and fermionic fields of the coset respectively) from the left, that is $g'=\Omega g(w,\theta)$. From the properties of $g_{OSP(6|4)}$ it follows that we can write
\be\label{cosettrans}
g'=g_{OSP(6|4)}(w',\theta')\,\omega(w,\theta)\,e^{\varphi \mathcal{\epsilon}}\,e^{v_{\lambda} \, Q^{\lambda}},
\ee
where $\omega$ is the compensating element from the stabilizer $U(3)$ of the $SO(6)/U(3)$ coset.
Let us write $\omega = \omega_{SU(3)} \omega_{U(1)}$, where $\omega_{SU(3)}$ belongs to $SU(3)\subset U(3)$ and therefore also to $SO(7)$, and $\omega_{U(1)}=e^{\nu K_6}\in U(1)\subset U(3)$ does not belong to $SO(7)$. Then $\omega_{U(1)} e^{\varphi \mathcal{\epsilon}}\,e^{v_{\lambda} \, Q^{\lambda}}=e^{\varphi \mathcal{\epsilon}}\,e^{v_{\lambda} \, Q^{\lambda}} \omega_{U(1)} = e^{(\varphi+3\nu) \mathcal{\epsilon}} (e^{-3\nu\epsilon} e^{v_{\lambda} \, Q^{\lambda}} e^{3\nu\epsilon}) e^{\nu U}$, where, as before, $U=K_6-3\epsilon \in SO(7)$ belongs to the stabilizer of the $SO(8)/SO(7)$ coset. We denote 
\bea
&\varphi'=\varphi+3\nu,& \\ &v_\lambda'Q^{\lambda}=e^{-3\nu\epsilon} v_\lambda Q^{\lambda} e^{3\nu\epsilon},&
\eea
then we may write
\be
g'=g_{OSP(6|4)}(w',\theta') \,e^{\varphi' \mathcal{\epsilon}}\,e^{v_{\lambda}' \, Q^{\lambda}}\,e^{\nu U}\,\omega_{SU(3)}.
\ee
The key property to observe is that, since $\nu$ is a function of $w, \bar{w}, \theta$, the variations of all fields involve only the fields $w, \bar{w}$ and $\theta, v$ (we have included $v$ here, since the variation of $v$ is $\delta v \sim \nu\, v$), but not $\varphi$. In order to see why this is important we write down the 11D metric in the Kaluza-Klein form (in the presence of fermions):
\be\label{kkm}
ds^{2}=ds^2_{10D}+\widehat{\mathcal{D}}\,(d\varphi-\mathcal{A}^{(1)})^2,
\ee
where $\widehat{\mathcal{D}}$ is the dilaton. Since the variations of all the fields do not depend on $\varphi$, the only way for (\ref{kkm}) to be invariant is for both $d\varphi-\mathcal{A}^{(1)}$ and $\widehat{\mathcal{D}}$ to be invariant.

\section{Low-energy limit: integrating out the massive fields}\label{low-energyapp}

Usually the low-energy limit implies that we need to get rid in one or another way of all the massive fields in the theory. As explained in Section \ref{low-energy}, in the present model finding the low-energy limit is a rather subtle endeavour. This is due to the fact, that the Lagrangian contains terms of various dimensions. Here by dimension we always mean the simple canonical dimension, which in turn is determined by the behaviour of the propagator of a given field: all scalar fields have dimension 0 and fermions have dimension 1/2, the derivative being clearly of dimension one.  Let us make it clear that so far we have only fixed the conformal gauge $\gamma_{+-}=\gamma_{-+}=1$ and set to zero the massive $\theta$ fermions. Thus, we should keep in mind that we have the v-fermions, some of which are redundant and are subject to the additional kappa-symmetry transformations and, last but not least, some of the bosonic fields in the Lagrangian (there are 10 bosonic fields) are unphysical and are subject to the Virasoro constraints. Ultimately we want to express the two gauge artifacts $z_\pm$ in terms of the physical fields. Of course, this is best done using the light-cone gauge \cite{uv1}, but for our purposes it will be enough to use a shortcut which we will now describe. The Virasoro constraints look as follows:
\bea\label{Vir1s}
0=V_1 = \del_+ z_+ - z_1 +{i\over 2} (\xi_1 \xi_2+\phi_1 \phi_2 - \chi_1 \chi_2 - \psi_1 \psi_2)+...\\ \label{Vir2s}
0=V_2 = \del_- z_- + z_1 + {i\over 2} (\xi_1 \xi_2+\phi_1 \phi_2 - \chi_1 \chi_2 - \psi_1 \psi_2)+... ,
\eea
The part of the Lagrangian, which contains the $z$ fields, is:
\bea \label{lagrz}
&\mathcal{L}_z= 2 z_2^2 + 4 z_1 \del_+ z_+ - 4 z_1 \del_- z_- +&\\ \nonumber &+ 4  \del_- z_-\del_+ z_+ + 2i (\del_- z_-+\del_+ z_+) (\xi_1 \xi_2+\phi_1 \phi_2-\psi_1 \psi_2 - \chi_1\chi_2)+&\\ \nonumber&+4i z_1 (\xi_1 \psi_2+\xi_2 \psi_1-\phi_1 \chi_2-\phi_2 \chi_1)+2i z_2 (\chi_2 \psi_1 +\chi_1 \psi_2-\xi_1 \phi_2-\xi_2 \phi_1)&
\eea
We remind the reader that the physical fields $z_1, z_2$ are massive, so we may set them to zero everywhere, except for the terms written above (since some of these terms have a 'subcritical' dimension, that is dimension smaller than 2). For this reason we did not write out the kinetic terms of the $z_1, z_2$ fields above (we're going to integrate out the $z_1, z_2$ fields, in the same fashion as the $W$ and $Z$ bosons can be integrated out in the low-energy limit of the Standard Model).

Next we plug $\del_+ z_+,  \del_- z_-$ from (\ref{Vir1s}, \ref{Vir2s}) into (\ref{lagrz}) to obtain:
\bea\label{lz0}
&\mathcal{L}_z =4 z_1^2 +2 z_2^2 +&\\ \nonumber &+4i z_1 (\xi_1 \psi_2+\xi_2 \psi_1-\phi_1 \chi_2-\phi_2 \chi_1)+2i z_2 (\chi_2 \psi_1 +\chi_1 \psi_2-\xi_1 \phi_2-\xi_2 \phi_1)+&\\ \nonumber &+ 2 (\xi_1 \xi_2 \phi_1 \phi_2 - \xi_1 \xi_2 \chi_1 \chi_2-\xi_1\xi_2 \psi_1 \psi_2-\phi_1\phi_2 \chi_1 \chi_2 -\phi_1 \phi_2 \psi_1 \psi_2 +\chi_1\chi_2 \psi_1 \psi_2)&
\eea
As an intermediate result, we get the correct masses (2 and 4) for the AdS physical fields (this should be compared with the spectrum obtained for the first time in \cite{ft}). We can now easily integrate out the fields $z_1$ and $z_2$:
\bea\label{lz}
\mathcal{L}_z= 3 \xi_1 \xi_2 \phi_1 \phi_2-2 \xi_1\xi_2
  \chi_1 \chi_2-2\xi_1\phi_1\chi_2\psi_2-3
  \xi_1 \phi_2\chi_1\psi_2-\\ \nonumber-\xi_1\phi_2
   \chi_2 \psi_1-\xi_2\phi_1\chi_1\psi_2-3
   \xi_2 \phi_1\chi_2\psi_1-2 \xi_2\phi_2
   \chi_1\psi_1-\\ \nonumber -2 \phi_1\phi_2\psi_1\psi_2+3
   \chi_1\chi_2\psi_1\psi_2
\eea
Recall that (\ref{lz}) is only the part of the Lagrangian which initially depended on the $z$ fields. There's another part, which we would have obtained, had we simply set all the $z$ fields to zero and dropped the vertices with higher derivatives. This part looks as follows:
\bea \label{l0}
&\mathcal{L}_0 =
2 \,(\frac{\delta_{i j}}{1+|w_{k}|^2}-\frac{\bar{w}_{i}\,w_{j}}{(1+|w_{k}|^2)^2}) \,(\partial_+ w_{i} \partial_-\bar{w}_j+\partial_- w_{i} \partial_+ \bar{w}_j)+&\\\nonumber &+
i (\xi_1 - \psi_1) \del_+ (\xi_1-\psi_1)+i (\phi_1 + \chi_1) \del_+ (\phi_1 + \chi_1)+&\\ \nonumber
&- i (\xi_2 - \psi_2) \del_- (\xi_2-\psi_2)-i (\phi_2 + \chi_2) \del_- (\phi_2 + \chi_2)+&\\ \nonumber &+
2 \, \frac{\bar{w}_i \del_+ w_i - w_i \del_+ \bar{w}_i}{1+|w_k|^2}\, (\xi_1-\psi_1) (\phi_1+\chi_1)-2 \, \frac{\bar{w}_i \del_- w_i - w_i \del_- \bar{w}_i}{1+|w_k|^2}\, (\xi_2-\psi_2) (\phi_2+\chi_2)+&
\\ \nonumber
&+3 (\xi_1 \xi_2 \phi_1 \chi_2  -   \xi_1 \xi_2 \phi_2 \chi_1   -  \xi_1 \phi_1 \phi_2 \psi_2  -  \xi_1 \chi_1 \chi_2 \psi_2 +&\\ \nonumber &+ \xi_2 \phi_1 \phi_2 \psi_1+\xi_2 \chi_1 \chi_2 \psi_1 + \phi_1 \chi_2 \psi_1 \psi_2 - \phi_2 \chi_1 \psi_1 \psi_2)+&\\ \nonumber &+5(\xi_1 \xi_2\chi_1 \chi_2 + \phi_1 \phi_2 \psi_1 \psi_2)+6(\xi_1 \phi_2 \chi_1 \psi_2 + \xi_2 \phi_1 \chi_2 \psi_1)-&\\ \nonumber &-\xi_1 \phi_1 \chi_2 \psi_2+\xi_1 \phi_2 \chi_2 \psi_1 + \xi_2 \phi_1 \chi_1 \psi_2-\psi_2 \phi_2 \chi_1 \psi_1&
\eea
The indices $i, j, k$ in this formula run over the values $1,2,3$. As explained in the body of the paper, the second and third lines of (\ref{l0}) determine the kappa-symmetry transformations, and it is clear, that the quartic terms in this formula are not invariant under this transformation (indeed, the only invariant combination is $(\xi_1 - \psi_1) (\xi_2 - \psi_2) (\phi_1 + \chi_1) (\phi_2 + \chi_2) $). However, according to the general logic that we have explained, the complete low-energy Lagrangian is the sum of (\ref{lz}) and (\ref{l0}):
\bea
&\mathcal{L} = 
2 \,(\frac{\delta_{i j}}{1+|w_{k}|^2}-\frac{\bar{w}_{i}\,w_{j}}{(1+|w_{k}|^2)^2}) \,(\partial_+ w_{i} \partial_-\bar{w}_j+\partial_- w_{i} \partial_+ \bar{w}_j)+&\\ \nonumber&+i (\xi_1 - \psi_1) \del_+ (\xi_1-\psi_1)+i (\phi_1 + \chi_1) \del_+ (\phi_1 + \chi_1)+&
\\ \nonumber &- i (\xi_2 - \psi_2) \del_- (\xi_2-\psi_2)-i (\phi_2 + \chi_2) \del_- (\phi_2 + \chi_2)+&
\\ \nonumber &+
2 \frac{\bar{w}_i \del_+ w_i - w_i \del_+ \bar{w}_i}{1+w_i^2} (\xi_1-\psi_1) (\phi_1+\chi_1)-2 \frac{\bar{w}_i \del_- w_i - w_i \del_- \bar{w}_i}{1+w_i^2} (\xi_2-\psi_2) (\phi_2+\chi_2)+&\\ \nonumber
&+
3 (\xi_1 - \psi_1) (\xi_2 - \psi_2) (\phi_1 + \chi_1) (\phi_2 + \chi_2)&
\eea
Thus, the result is invariant under the residual kappa transformations, as it should be. Now we can safely set to zero 4 of the 8 fermions to obtain the Lagrangian announced in the body of the paper. Note that in order to obtain (\ref{lagr2}) or (\ref{lagr3}), one needs to rescale the fermions by a factor of 2 ($v\to 2\, v$) and divide the Lagrangian by 4.

\section{Matrices and notations}\label{matrices}

Here we explicitly present the matrices, which appeared in the main text.

{\bf AdS $\Gamma$-matrices}

The representation of the four $\Gamma$ matrices (which come from the $AdS$ space) used throughout the paper is as follows:
\begin{equation}\nonumber \small
\Gamma_{0}= \begin{pmatrix}
1 &0&0&0\\
0&1&0&0\\
0&0&-1&0\\
0&0&0&-1
\end{pmatrix},\;
\Gamma_{1}= \begin{pmatrix}
0&0&0&1\\
0&0&1&0\\
0&-1&0&0\\
-1&0&0&0
\end{pmatrix},\;
\Gamma_{2}= \begin{pmatrix}
0&0&0&-i\\
0&0&i&0\\
0&i&0&0\\
-i&0&0&0
\end{pmatrix},
\Gamma_{3}= \begin{pmatrix}
0 &0&1&0\\
0&0&0&-1\\
-1&0&0&0\\
0&1&0&0
\end{pmatrix},\;
\end{equation}
One can observe that for $k=1,2,3$ we have $\Gamma_{k}=i\sigma_{2} \otimes \sigma_{k}$. Besides, we introduce $\Gamma_5=i \Gamma_0 \Gamma_1 \Gamma_2 \Gamma_3$. Another matrix encountered in the text is
\begin{equation}\nonumber
C_{4}=i\Gamma_{0}\Gamma_{2}=\begin{pmatrix}
0 &0&0&1\\
0&0&-1&0\\
0&1&0&0\\
-1&0&0&0
\end{pmatrix}
\end{equation}

{\bf $\mathbb{C}P$ $\gamma$-matrices}

The seven $\gamma$ matrices (which come from the $S^7$ space) are as follows:
\bea\nonumber
&\gamma_1 = I_2 \otimes \Gamma_3,\;\gamma_2=\sigma_3\otimes\Gamma_1,\;\gamma_3=i\sigma_2 \otimes i \Gamma_0\Gamma_1\Gamma_2 &\\
\nonumber &\gamma_4=i\sigma_2 \otimes i\Gamma_2\Gamma_3,\;\gamma_5=i\sigma_2 \otimes \Gamma_0 \Gamma_3,\;\gamma_6=-\sigma_1\otimes \Gamma_1,\; \gamma_7=I_2 \otimes \Gamma_3 \Gamma_1,&
\eea
The following matrices were also encountered in the text:
\bea
&\nonumber U={\rm diag}(-1,-1,-1,3)\otimes i\sigma_2,&\\
&\nonumber K_6={\rm diag}(1,1,1,0)\otimes i\sigma_2,&\\
&\nonumber \mathcal{\epsilon}={\rm diag}(0,0,0,1)\otimes i\sigma_2&
\eea

{\bf 11D $\mathbf{\Gamma}$-matrices}

The 11D $\mathbf{\Gamma}$ matrices were used in formula (\ref{WZmembrane}) to construct the Wess-Zumino term of the M2 brane action. These matrices (which have dimensionality 32) can be built in the following way, as tensor products of the $AdS$ and $\mathbb{C}P$ gamma-matrices defined above:
\bea\nonumber
\mathbf{\Gamma}^{A}=\{ \,i\Gamma^a \Gamma^5 \otimes I_8,\,a=0 ... 3; \;\Gamma^5 \otimes \gamma^b,\,b=1 ... 7\,\}
\eea
The 11D charge-conjugation matrix looks as follows:
\be
C_{11} = C_4 \Gamma_5 \otimes I_8 
\ee
and has the property
\be
(\mathbf{\Gamma}^{A})^{{\rm T}} = - C_{11} \mathbf{\Gamma}^{A} C_{11}^{-1} 
\ee

{\bf The "v-fermions"}

The $v_\lambda Q^\lambda$ matrix of v-fermions explicitly looks as follows:

\be\label{vferm}
v = v_\lambda Q^\lambda =\begin{bmatrix}
0&0&0&0&0&0&v_{17}&v_{18}\\
0&0&0&0&0&0&v_{27}&v_{28}\\
0&0&0&0&0&0&-v_{27}^\ast&-v_{28}^\ast\\
0&0&0&0&0&0&v_{17}^\ast&v_{18}^\ast \\
\end{bmatrix},
\ee
where
\bea
v_{17}= \frac{1}{\sqrt{2}} e^{-i \frac{\pi}{4}}\,(\psi_{1}+i\, \psi_{2}),\;\;v_{18}=\frac{1}{\sqrt{2}} e^{-i \frac{\pi}{4}}\,(\chi_{1}+i \,\chi_{2}),\\
v_{27}=, \frac{1}{\sqrt{2}} e^{-i \frac{\pi}{4}}\,(\phi_{1}+i\, \phi_{2})\;\;v_{28}=  \frac{1}{\sqrt{2}} e^{-i \frac{\pi}{4}}\,(\xi_{1}+i\, \xi_{2}).
\eea
On several occasions we used the complex fermion notation $\chi=\psi_1+i\chi_1,\;\psi=\psi_2+i\chi_2,$ $\xi=\xi_1 - i \phi_1 ,\; \eta=\xi_2 -i \phi_2$. Besides, in the paper we made use of the gauge (\ref{kappagauge}), which stands for $\phi_1=\phi_2=\xi_1=\xi_2=0$.

\end{document}